\documentclass[iop]{emulateapj}
\shorttitle{Giant Molecular Clouds in IC 342}
\shortauthors{Hirota et al.}
\begin{document}
\title{GIANT MOLECULAR CLOUDS IN THE SPIRAL ARM OF IC 342}
\author{
    Akihiko \textsc{Hirota}, \altaffilmark{1}
    Nario \textsc{Kuno}, \altaffilmark{1,2}
    Naoko \textsc{Sato}, \altaffilmark{3} 
    Hiroyuki \textsc{Nakanishi}, \altaffilmark{4}  
    Tomoka \textsc{Tosaki}, \altaffilmark{1,6}
    and 
    Kazuo \textsc{Sorai} \altaffilmark{5}
} 
\altaffiltext{1}{Nobeyama Radio Observatory, Minamimaki, Minamisaku, Nagano 384-1805, Japan} 
\altaffiltext{2}{The Graduate University of Advanced Studies (SOKENDAI), 2-21-1 Osawa, Mitaka, Tokyo 181-8588, Japan}
\altaffiltext{3}{Student Center for Independent Research in the Science, Wakayama University, Wakayama-shi, Wakayama 640-8510, Japan} 
\altaffiltext{4}{Graduate School of Science and Engineering, Kagoshima University,1-21-35 Korimoto, Kagoshima, Kagoshima 890-0065, Japan}
\altaffiltext{5}{Department of Physics / Department of Cosmosciences, Hokkaido University, Kita-ku, Sapporo 060-0810}
\altaffiltext{6}{Current address: Joetsu University of Education, 1 Yamayashiki-machi, Joetsu, Niigata, 943-8512, Japan}
% Abstract
%----------------------------------------
\begin{abstract}
We present results of $^{12}$CO (1--0) and $^{13}$CO (1--0) observations
of the northeastern spiral arm segment of \objectname{IC 342} with a $\sim$ 50pc resolution
carried out with the Nobeyama Millimeter Array.
Zero-spacing components were recovered by combining with the existing data taken with the Nobeyama 45m telescope.
The objective of this study is to investigate the variation of cloud properties across the spiral arm with a resolution comparable to the size of giant molecular clouds (GMCs).
The observations cover a 1 kpc $\times$ 1.5 kpc region located $\sim$ 2 kpc away from the galactic center, where a giant molecular association is located at trailing side and 
associated star forming regions at leading side.
\ \ The spiral arm segment was resolved into a number of clouds whose 
size, temperature and surface mass density are comparable to typical GMCs in the Galaxy. 
Twenty-six clouds were identified from the combined data cube and 
the identified clouds followed the line width-size relation of the Galactic GMCs.
The identified GMCs were divided into two categories according to whether 
they are associated with star formation activity or not.
Comparison between both categories indicated 
that the active GMCs are more 
massive, have smaller line width, and are closer to virial equilibrium compared to the 
quiescent GMCs.
These variations of the GMC properties suggest that dissipation of excess kinetic energy of GMC 
is a required condition for the onset of massive star formation.
%   Comparison with star formation tracers indicated that 
%   while some of the GMCs are closely associated with star forming regions, 
%   the rest of the GMCs show little or no sign of massive star formation. 
%
%
\end{abstract}
\keywords{galaxies: individual (IC342) - galaxies: ISM - galaxies:spiral - ISM:clouds- ISM:molecules }
\section{INTRODUCTION}
\ \ The formation of massive star is one of the fundamental processes 
driving secular evolution of spiral galaxies.
The rate of massive star formation in galaxies is 
considered to be regulated by 
the formation and the evolution of giant molecular clouds (GMCs),
which are progenitors of massive stars.
However, the exact processes which drive the formation and the evolution of GMC to 
initiate massive star formation is still unclear.\\
\ \ GMCs are large cloud complexes with mass of 10$^5$ to several times of 10$^6$ $M_{\odot}$ and 
size of 20-100pc \citep[e.g.,][]{ SandersScovilleSolomon1985, Solomon1987Larson, 
Dame1986FirstQuadrant}.
GMCs comprise roughly $80\%$ of the total molecular mass in the inner Galaxy 
\citep{SandersScovilleSolomon1985, Solomon1987Larson}.
Although the mean volume density of GMCs is as low as $\sim$ 100 cm$^{-3}$, 
there is strong density contrasts inside the GMCs, 
which make local volume density higher than $10^5$ cm$^{-3}$, 
which is thought as critical density for the onset of the star formation 
\citep[e.g.,][]{Elmegreen2002SFOverCriticalDensity}.
As formation of dense clump is essential in forming stars, 
interpretation of the physical process which make a density contrast inside GMC is crucial.\\
%Those dense and clumpy structures seen inside GMCs are called as 'cloud core'.
%Formation of 'cloud core' is essential in forming stars.
\ \ Strong scaling relations between the size and line width of GMCs were found both in the Galactic clouds
\citep[e.g.,][]{Larson1981, Dame1986FirstQuadrant, Solomon1987Larson, Heyer2004TurbulenceMC} 
and in the extragalactic clouds \citep[e.g.,][]{Rosolowsky2003M33, Bolatto2003SMC}.
The scaling relation holds not only across clouds but also within clouds,
suggesting that entire structures of GMCs are strongly governed by supersonic turbulent motion
\citep[e.g,][]{Larson1981, Heyer2004TurbulenceMC}.
%--------------------
% before correction
%--------------------
%GMCs comprise roughly $80\%$ of the total molecular mass in the inner Galaxy and are highly structured \citep{SandersScovilleSolomon1985, Solomon1987Larson}.
%------------------------
%
% Heyer 200 \& Brunt 2004 -> Larson's law extends within individual GMCs
The supersonic turbulence act in two ways: while it supports the GMCs against 
global gravitational collapse,
it also plays an important role in forming density structures within 
each GMC \citep[{\it turbulent fragmentation}, e.g., ][]{VazquezSemadeni1994, Padoan2002TurbulentFragmentation2CloudCore}.
There is an argument that, as a consequence of the fragmentation process, 
dense cloud cores, which finally collapse into stars, are formed inside GMCs
\citep[e.g.,][]{Elmegreen2002SFOverCriticalDensity, MacLow2004TurbulenceRegulatedSF, Krumholz2005TurbulenceSF}, 
although additional process, such as competitive accretion
\citep[][]{Bonnell2001a,BonnellBate2006ThroughGravAndCA}, 
might be also required in massive star formation.
An investigation of the dynamical properties of GMCs is important in discussing 
the formation and evolution processes of GMCs. \\
\ \ Recent molecular cloud studies in the Galaxy indicated that 
GMCs which lack associated star formation activity are not 
as unusual as once thought
\citep[e.g.,][]{Chiar1994CO21Survey}. 
A comparison of cloud properties between star-forming GMCs and quiescent GMCs 
is important in addressing 
what really controls the star formation activity in GMCs. 
\citet{Williams1994CLFIND} compared cloud properties between 
a quiescent cloud \citep[\objectname{G216-2.6}, known as \objectname{Maddalena's cloud},][]{MaddalenaThaddeus1985}
and an active cloud \citep[the \objectname{Rosette} molecular cloud,][]{BlitzThaddeus1980Rosette}. 
They showed that the line width is larger in the Maddalena's cloud than in the \objectname{Rosette} cloud
over every size scale. 
However, distance degeneracy in observations of Galactic clouds prevents further 
investigation of the relation between cloud properties and the star formation activity. 
To overcome this limitation, 
observations of molecular clouds in external galaxies are crucial.\\
\ \ Early millimeter molecular-line observations in the external grand-design spiral galaxies often found large associations of molecular clouds
\citep[e.g.,][]{VogelKulkarniScoville1988, Rand1995AJNGC4321, RandLordHidgon1999M83}. 
%
%  ここら辺り変えた、少し見直しが必要
% 
Mass of those clouds were larger than GMCs by over an order of magnitude 
(greater than several times of $10^7 M_{\odot}$) 
and termed as giant molecular association (GMA) 
\citep{VogelKulkarniScoville1988}. 
Distributions of those GMAs were found to be coherently offset from 
associated H$_{\rm{II}}$ regions and interpreted as 
representative of 
the time delay between the formation of those massive clouds and the onset of 
the subsequent massive star formation
\citep[e.g.,][]{Rand1993M51CO}.
By resolving the spatial offsets down to the scale of GMCs, 
it should be able to investigate the evolution sequence of GMCs which leads to the massive star formation. \\
\ \ We report the results of the observations 
of the nearby galaxy \objectname{IC 342} carried out with the 
Nobeyama Millimeter Array (NMA) in $^{12}$CO (1--0) and $^{13}$CO (1--0) lines.
The observed field covers a 1 kpc $\times$ 1.5 kpc region located in the northeastern spiral arm segments, 2kpc away from the galactic center. 
At the site, spatial offsets between molecular clouds and 
star forming regions were previously identified by the Nobeyama 45m telescope  
observation \citep{Hirota2010IC34213CO}.
The prime target of this study is to investigate variation of cloud properties across a 
spiral arm in \objectname{IC 342} with a spatial resolution comparable to the size of GMCs
\citep[$\sim$ 50pc;][]{SandersScovilleSolomon1985, Blitz1993MolecularClouds}.\\
\ \ \objectname{IC 342} is a nearby spiral galaxy classified as SAB(rs)cd \citep{deVaucouleurs1991RC3}. 
%Because of the log galactic latitude of the galaxy $l \sim 10.6^{\circ}$, 
%morphology of the galaxy is not clear at the first glance. 
The central region of \objectname{IC 342} harbors starburst activity \citep[e.g.,][]{Becklin1980IC342IR} and because of the strong millimeter and submillimeter emission from the region, 
the central region (within $\sim$ 500pc) of the galaxy has been studied in detail.
The distributions of molecular clouds is indicative of bar potential
\citep[e.g.,][]{Ishizuki1990IC342, TurnerHurt1992ic342co13}, and
molecular gases are likely to be excited by starburst activity 
\citep[e.g.,][]{WallJaffe1990, Eckart1990IC342, Turner1993ic342co2to1, Schulz2001IC342, Meier2000IC34213CO21, IsraelBaas2003ic342maf2}.
Studies of molecular chemistry indicated the chemistry of the clouds in the central region are also subject to shocks \citep[][]{MeierTurner2005, Usero2006IC342} presumably driven by the bar kinematics. 
Supply of material driven by the bar kinematics is likely to be regulated by negative mechanical feedback 
from the starburst activity \citep{Schinnerer2008IC342}.
Beyond the radius of 500pc, the distribution of the molecular gas disk of \objectname{IC 342} was studied with the single dish telescope observation mostly in $^{12}$CO (1--0)
\citep{SageSolomon1991ApJ342, Crosthwaite2001, Sato2006, Kuno2007Atlas}.
In particular, the $^{12}$CO (1--0) map presented in \citet{Kuno2007Atlas} reveal 
prominent galactic structures such as a bar and spiral arms.
These structures are also visible in near-infrared images \citep{Jarrett2003}, suggesting the 
existence of moderately strong density waves in the galaxy.
The proximity \citep[3.3 Mpc,][]{SahaClaverHoessel2002} of the galaxy provides an
opportunity to investigate the influence of density waves exerted on molecular clouds.
\section {OBSERVATION AND DATA REDUCTION}
\subsection {$^{12}$CO (1--0) and $^{13}$CO (1--0) observation}
\begin{figure} [htbp]
  \begin {center}
    %\plotone{img/45m_guide_img.eps}
    \plotone{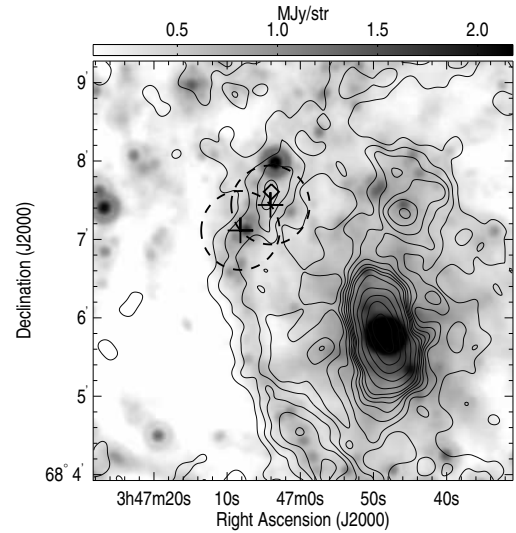}
%	\caption {$^{12}$CO (1--0) integrated intensity map of the northeastern arm segments of IC 342. The contour levels are 1, 2, 3,..., 7 times 5 K km s$^{-1}$. The dotted circles represents the primary beam patterns}
	\caption {Velocity-integrated image of $^{12}$CO (1--0) data taken from \citet{Sato2006} (contour) compared with 24$\mu$m image taken with the Multiband Imaging Photometer for {\it Spitzer} \citep[MIPS;][]{Rieke2004MIPS} camera.
    Contour levels are 10, 15, 20, 25, 30, 35, 40, 50, 70, 100, 140, and 190 K km s$^{-1}$. 
    Pointing centers of the NMA observations are indicated with crosses.
    Dashed circles with size of 60${''}$ indicate approximate size of the
    NMA field of view ($\sim 59{''}$ for $^{12}$CO (1--0) and $\sim 62{''}$ for $^{13}$CO (1--0)).
    The peak position of the GMA seen in the $^{12}$CO image is indicated with a diamond.
}
	\label {fig: 45m_guide_img}
  \end {center}
\end{figure}

\ \ Aperture synthesis observations of the northeastern spiral arm segments of \objectname{IC 342} in $^{12}$CO(1--0) and $^{13}$CO (1--0) were carried out with the NMA. 
The NMA consists of six 10m antennas which provide field-of-view size of 
$\sim$ 59$''$ and $\sim$ 62${''}$ at the rest frequency of $^{12}$CO (1--0) and $^{13}$CO (1--0),
respectively.
 Three antenna configurations (AB+C+D) were used to sample visibility in the $uv$ range of $\sim$ 4 $k\lambda$ 
to $\sim$ 130 $k\lambda$.
The observed region covers a GMA located on the northeastern spiral arm segments, 
where spatial offsets among $^{12}$CO (1--0), $^{13}$CO (1--0) and star formation tracers exist \citep{Hirota2010IC34213CO}.
To cover the region of interest, two pointings separated from each other by 30${''}$ were set (Figure \ref{fig: 45m_guide_img}). 
%----------------------------------
% Figure - Guide Image
%----------------------------------
\ \ The observations were carried out from 2005 November to 2007 April 
for $^{12}$CO (1--0) and from 2006 November to 2007 March for $^{13}$CO (1--0). 
%
%We made a two points mosaicing observation to cover the region of interest.  
System noise temperatures (in single side band) were 500-1000K for $^{12}$CO (1--0) observations and 400-700K for $^{13}$CO (1--0) observations.
The Ultra-Wide-Band-Correlator \citep[][]{Okumura2000UWBC} configured to cover 256MHz bandwidth with 256 channels was used as the backend.
Window function for the spectrometer was set to Hanning smooth function and the resultant frequency resolution was 2MHz. 
Two pointings were observed alternately every 8 minutes to attain a uniform {\it uv}-coverage. 
B0355+508 was observed once in every $\sim$ 20 minutes as a gain calibrator and 
3C273 was observed to determine the pass-band.\\
\ \ Acquired raw visibility data were calibrated using the software package UVPROC-II \citep{Tsutsumi1997UVPROC2}. 
Though the NMA data were combined with the data observed with the 45m telescope afterward, 
map image of the CO lines were first made without combining single-dish data to check the 
basic quality of the data.
Imaging and deconvolution were made following the standard procedures implemented in the software package MIRIAD \citep{SaultTeubenWright1995MIRIAD}.
Resolution of the resultant $^{12}$CO data cube was $3{''}.1 \times 2{''}.5$ in spatial directions and 5.2 km s$^{-1}$ in velocity direction.
Typical rms noise was $\sim$ 24 $\mathrm{mJy}$ beam$^{-1}$ within each channel. 
For $^{13}$CO (1--0) data cube, spatial and velocity resolutions were $3{''}.88 \times 3{''}.53$ and 

$\sim$ 5.4km s$^{-1}$, respectively, and typical rms noise was $\sim$ 11 $\mathrm{mJy}$ within each channel.
The total amount of the flux detected with the NMA is 
% Jy beam-1 km/s
%   $\sim 2.2\times 10^6$ Jy km s$^{-1}$ for $^{12}$CO (1--0) and 
%   $\sim 1.1\times 10^4$ Jy km s$^{-1}$ for $^{13}$CO (1--0), respectively.
%   While on the other hand, total flux emitted from the observed region is estimated to be 
%   $\sim 3.0\times 10^5$ Jy km s$^{-1}$ for $^{12}$CO (1--0) and 
%   $\sim 2.1\times 10^4$ Jy km s$^{-1}$ for $^{13}$CO (1--0) from the 45m data.
$\sim 8.7 \times 10^1$ Jy km s$^{-1}$ for $^{12}$CO (1--0) and 
$\sim 4.0 $ Jy km s$^{-1}$ for $^{13}$CO (1--0).
While, on the other hand, the total flux emitted from the observed region was estimated from 
the 45m data and was 
$\sim 6.1 \times 10^2$ Jy km s$^{-1}$ for $^{12}$CO (1--0) and 
$\sim 8.7 \times 10^1$ Jy km s$^{-1}$ for $^{13}$CO (1--0).
%
% 12CO(SD) -> 2450 Jy km/s 12CO(INT)-> 347 Jy km/s    -> Mistake!!
% 13CO(SD) ->  317 Jy km/s 13CO(INT) -> 16 Jy km/s    -> Mistake
% 12CO(SD) ->  616Jy km/s 12CO(INT)->  86.7Jy km/s    
% 13CO(SD) ->  79.2477 Jy km/s 13CO(INT) -> 4.013 Jy km/s
The large amount of the 'missing' flux discrepancies indicates 
the presence of extended structures whose spatial frequency is lower than $\sim$ 4.5 $k\lambda$, 
which is the minimum spatial frequency available with the NMA.
\subsection {NMA and 45m Combined Data: Short Spacing Correction}
\ \ To recover the missing extended flux, the NMA data were combined with the 
$uv$ data generated from the 45m data to fill in the central 'hole' in the $uv$-coverage.
$^{12}$CO (1--0) data were taken from \citet{Sato2006} and 
$^{13}$CO (1--0) data from \citet{Hirota2010IC34213CO}.
Combining procedures follow the method described in \citet{Takakuwa2003Combine} and \citet{Kurono2009}. 
We will denote outlines of the procedures hereafter.
The 45m data were deconvolved with a Gaussian pattern using a Wiener filter 
and multiplied with the NMA primary beam pattern to simulate the NMA observation.
Visibility data were generated from the 45m data by performing Fourier transformation on the deconvolved 45m data.
On generating visibility data, $(u,v)$-sampling points were uniformly distributed within the radius of $4.5$ k$\lambda$, 
which corresponds to the 'central hole' of the NMA $(u,v)$-coverage. 
Both the NMA and the 45m visibility data sets were merged to create a single visibility data set and imaged with the inverse Fourier transformation. 
"Clean" deconvolution was performed with the MOSSDI task implemented in the MIRIAD.
Finally, the images were corrected for primary-beam attenuation by dividing with the gain distribution estimated with the standard MIRIAD task (MOSSEN). \\
\ \ On performing the inverse Fourier transformation, 
weights attached to the 45m and the NMA visibility determine the shape of the synthesized beam. 
If larger weights are attached to the $45$m data, sensitivity to the low spatial frequency components in increased. 
However, if the 45m data are overweighted, the shape of the 
synthesized beam gets close to the $45$m beam and thus resolution gets worse.

In practice, the relative weights between the 45m data and the NMA data are determined by
the number of the $(u, v)$-sampling points generated from the 45m data 
and the integration time attached to each visibility sample for the 45m data.
An adequate choice of the parameters should attain the resolution comparable to the original NMA data and 
the complete recovery of the missing-flux \citep{Kurono2009}.
We generated several sets of visibility data for the $45$m data 
with different integration time attached to each visibility data sample.
The number of visibility data points were fixed since only the product of the both parameters is important.
Best weight parameter, which satisfies the requirement of the least amount of the missing flux and the least broadening of the beam, was selected from the trial data sets. \\
\ \ Properties of the final combined data cubes are listed in Table 
\ref{tab: data_parameters_combined}.
Spatial resolutions of the combined data cubes are almost identical to the original data cube.
The total flux within the observed fields are also almost identical to the single-dish data within the 
$\sim$ 5 \% level, which is well below calibration errors.
%The beam size of the resultant data is 3.2${''}$ $\times$ 2.58${''}$ and rms noise in each channel is $\sim$ 24 mJy beam$^{-1}$ km s$^{-1}$. 
%Figure \ref{fig: field_sp} shows the averaged spectrum of the NMA data, the 45m data, and the NMA \& 45m combined data inside the primary beam pattern of each pointing. \\
%
%
%
%
%\placetable {tab: data_params_combined}
\begin{deluxetable*}{ccccc}
    \tablecolumns{5}
    \tablewidth{0pt}
    \tabletypesize{\small}
    \tablecaption{Parameters of the Combined CO Data Cubes}
    \tablehead{
        \colhead{Line} &
        \colhead{Beam Size} &
        \colhead{Beam Position Angle} &
        \colhead{Velocity resolution} &
        \colhead{rms Noise}\\
        \colhead{} &
        \colhead{(arcsec)}  &
        \colhead{(deg)}  &
        \colhead{(km s$^{-1}$)} &
        \colhead{(mJy beam$^{-1}$)}
    }
    \startdata
        $^{12}$CO (1--0)& 3${''}$.20 $\times$ 2${''}$.58 & --72$^{\circ}$ & 5.2  & 23 \\
        $^{13}$CO (1--0)& 3${''}$.85 $\times$ 3${''}$.53 & --69$^{\circ}$ & 5.4  & 10 
    \enddata
    \label{tab: data_parameters_combined}
\end{deluxetable*}
\section {RESULTS}
%-----------------------------
% Distribution
%-----------------------------
\subsection {Molecular gas distribution: $^{12}$CO (1--0)}
\label{sec: 12co_distribution}
% NMA only
% NMA+45m
% NMA+45m+Halpha
% NMA+45m+(DSS)
% NMA+45m+(HI)
%\ \ Figrue. \ref{fig: a12ncomb_r5_m0}a) shows the integrated intensity image of the observed in 
%$^{12}$CO (1--0). The image is not combined with the 45m data and lacks for zero-spacing components.
%Total intensity image taken with the 45m telescope is shown in \ref{fig: a12ncomb_r5_m0}b) for comparison. 
%
\ \ The observed region contains a GMA which was visible with the previous CO observations
\citep{Kuno2007Atlas, Hirota2010IC34213CO}.
The NMA+45m combined data offer an opportunity to investigate the internal-structure of the GMA with a spatial resolution comparable to the size of GMCs. 
%
%
%----------------------------------------------
% NMA+45m combined image
%----------------------------------------------
\begin {figure*} [htbp]
    %\epsscale{1.7}
    %\plotone{img/m0_nma_combined.eps}
    \plotone{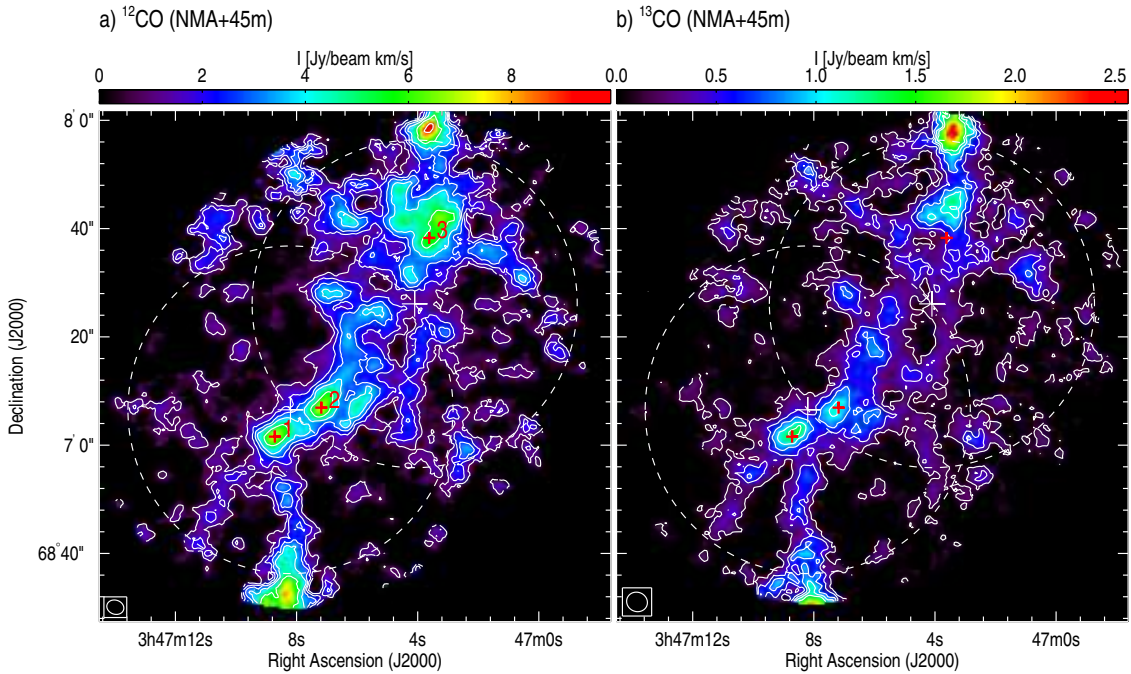}
    \caption {$^{12}$CO (1--0) and $^{13}$CO (1--0) integrated intensity images of the 
northeastern spiral arm segments of {IC 342}.
Locations of the three brightest $^{12}$CO sources are indicated with crosses (Points 1--3).
(a) $^{12}$CO (1--0) integrated intensity map of the NMA+ 45m combined data. 
Contour levels are 2, 4, 6, 8, 12, and 20 times 0.45 Jy beam$^{-1}$km s$^{-1}$. 
Dashed circles indicate the approximate size of the NMA field of view ($\sim 60{''}$).
The attenuation due to the primary beam pattern is corrected. 
(b) Same as (a), but for the $^{13}$CO (1--0) data. 
Contour levels are 1, 3, 5, 7, and 9 times 0.15 Jy beam$^{-1}$km s$^{-1}$. 
%   levs = 0.45 * [2, 4, 6, 8, 12, 20]
%            %levs=0.15*[1,3,5,7,9]
}
% scripts/batch_fig_m0_nma_combined.pro
    \label {fig: m0_nma_combined}
\end {figure*}
Figure \ref {fig: m0_nma_combined} shows the velocity-integrated intensity image of the combined $^{12}$CO(1--0) data. 
Several discrete sources with sizes of 50--100pc were found. 
For convenience, three brightest $^{12}$CO sources were marked with crosses in 
the figure (Points 1--3).
Indices were allocated in order of the $^{12}$CO intensity. 
Typical brightness temperatures ($T_{\rm{B}}$) at the peaks are 2--4 K above 
the cosmic background temperature, with a maximum temperature of 5.2 K at Point 1. 
Although these values are lower than the peak CO temperatures found in Galactic GMCs ($\sim$ 20K), 
it is comparable with the mean temperature averaged over full extent of cloud.
For example, 
\citet{Sakamoto1994Orion2to1} derived the spectrum of $^{12}$CO (1--0) and $^{12}$CO (2--1) averaged over almost entire extent of the Orion A and B clouds and
presented $^{12}$CO (2--1) peak intensities of 1.5K and 2.1K, and
CO (2--1) / CO (1--0) ratio of 0.75 and 0.62.
These values correspond to $^{12}$CO (1--0) temperature of about 2.1K and 3.4K for both the clouds.
Comparable brightness temperatures found with the NMA+45m data suggests that 
clouds seen with our observations may have similar nature with the Galactic clouds.\\
\ \ The characteristics of $^{12}$CO distribution on both the southern and the northern sides of the observed fields differ apparently: 
while it is concentrated in the narrow ridge structure on the southern side, 
it is more extended on the northern side. 
%------------------
% Southern half
%------------------
On the southern side, the molecular ridge consists of several discrete sources including Points 1 and 2.
%------------------
% Northern side
%------------------
In a contrasting situation, 
on the northern side, the cloud distributions are more extended compared to the southern side.
The GMA seen as a single peak in the previous 45m image 
was resolved into several minor peaks around Point 3.  \\
%
%
%
%-------------------------
% Molecular mass
%-------------------------
\ \ The amount of molecular gas mass is estimated from the CO integrated intensity 
by applying a "standard" CO--H$_2$ conversion factor of 
$X_{\rm{CO}}$ = 2 $\times$ 10$^{20}$ cm$^{-2}$ (K km s$^{-1}$)$^{-1}$ 
\citep[][]{StrongMattox1996, Dame2001}.
The surface mass density of H$_{2}$ is calculated by
\begin {equation}
    \Bigl(\frac{\Sigma_{\rm{H}_2}}{{M}_{\odot}\ \rm{pc}^{-2}}\Bigr) = 
        3.21\times\cos{i}\times\Bigl(\frac{I_{\rm{CO}}}{\rm{K\ km\ s}^{-1}}\Bigr),
    %\frac{I_\rm{CO}}{}
    %K\ km\ s^{-1}}
\end {equation}
where $i$ is the inclination angle of the galaxy \citep[$i=31^{\circ}$; ][]{Crosthwaite2000IC342HI}. 
%--------------
% He Correction
%--------------
The surface mass density of molecular gas corrected for He and other heavy elements is derived by
\begin {equation}
    {\Sigma_{\rm{mol}}} = 1.36\times\Sigma_{\rm{H}_2}.
\end {equation}
Highest peak surface molecular gas mass density within the observed region is 
$\sim 230 M_{\odot}$ pc$^{-2}$ found at Point 1. 
%----------------
% GMC surface mass
%----------------
Typical surface mass densities in other local peaks are $100--200 M_{\odot}$ pc$^{-2}$. 
These are consistent with the value ($\sim 100 M_{\odot}$ pc$^{-2}$) 
found in the Galactic clouds 
\citep[e.g.,][]{Solomon1987Larson, Blitz1993MolecularClouds}
and several nearby galaxies \citep[e.g.,][]{Rosolowsky2003M33,RosolowskyBlitz2005}.
The size, the temperature and the surface mass density of the discrete sources found in the $^{12}$CO map 
imply that the NMA+45m observations successfully resolved the molecular spiral arm 
into individual GMCs. 
%
%
%   \begin {figure} [htbp]
%     \begin {center}
%   	\plotone{img/a12ncomb_r5_m0.eps}
%   	\caption {$^{12}$CO (1--0) integrated intensity map of the NMA+ 45m combined data. Contour levels are 1, 2, 3,..., 9 times 0.8 Jy beam$^{-1}$km s$^{-1}$. The dotted circles represents the primary beam patterns. The attenuation due to the primary beam pattern is corrected. }
%   	\label {fig: a12ncomb_r5_m0}
%     \end {center}
%   % /home/hiko/temp/temp.pro
%   \end {figure}
%
%
%----------------------------------------------
% 13CO distribution
%----------------------------------------------
\subsection {Molecular gas distribution: $^{13}$CO (1--0)}
\label{sec: 13co_distribution}
%--------------------------------------------------
% unmasked
%
%   \ \ Figure \ref {fig: 13co_comb_m0_unmasked} is the velocity integrated intensity image 
%   of the $^{13}$CO(1--0) combined NMA + 45m data.
%   The emission below $2\sigma$ noise level are clipped. 
%   The peak brightness temperature is 1K above the CMB at Point 1. 
%--------------------------------------------------
\ \ Figure \ref{fig: m0_nma_combined}(b) shows an integrated intensity map of the combined $^{13}$CO (1--0) data. 
To improve the signal-to-noise ratio (S/N) of the integrated image, 
$^{13}$CO data were masked before calculating the integrated intensity.
The mask data were made according to the following procedures.
First, the $^{12}$CO data cube was smoothed and re-sampled to share the 
same resolution and pixel coordinates with the $^{13}$CO data cube.
Mask data was made from the smoothed $^{12}$CO data cube according to the following two-step procedures \citep{RosolowskyBlitz2005}:
first, pixels above 4$\sigma$ are adopted as 'kernel' mask and next, 
all the pixels above 2$\sigma$ and connected with the 'kernel' are included within the mask. 
Isolated masks smaller than the beam size were discarded. 
The applicapability of this mask to the $^{13}$CO data relies on the following issue:
as the critical volume density for the excitation of $^{12}$CO line 
($\sim 10^2$ cm$^{-3}$) is much lower than that of $^{13}$CO line ($\sim 10^3$ cm$^{-3}$), 
the extent of the $^{13}$CO emitting volumes should be enclosed within 
that of $^{12}$CO emission regions.
To check the validity of the mask, intensity histogram of $^{13}$CO 
for the residual pixels outside the mask was made. 
The shape of the residual $^{13}$CO intensity histogram was symmetric about zero and the central part of the histogram could be fitted with the Gaussian distribution with a sigma of $\sim$ 11 $\mathrm{mJy} \mathrm{beam}^{-1}$, which is consistent with the rms noise of the original $^{13}$CO data,
suggesting that almost all the $^{13}$CO emission were enclosed within the mask.
Finally, the mask was applied to the original $^{13}$CO data cube and integrated intensity was calculated.\\
%--------------------------------------
% 13CO integrated-intensity S/N ratio
%--------------------------------------
%The peak S/N ratio of the integrated intensity map of $^{13}$CO was 16 $\sigma$ at the 
%location of the strongest $^{12}$CO source.\\
\begin {figure*} [htbp]
    %\epsscale{2.0}
    %\plotone{img/a12_multic.eps}
    \plotone{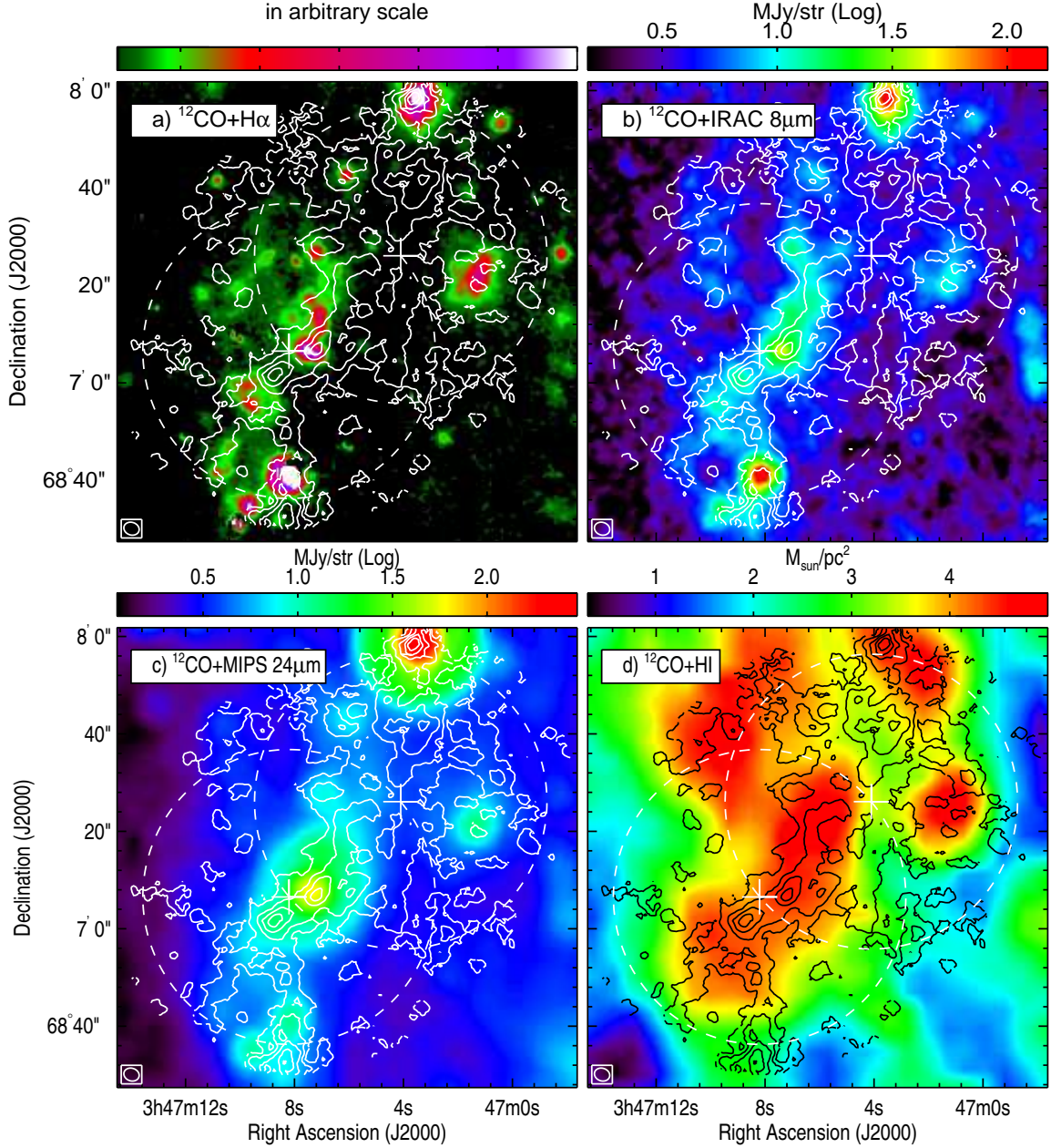}
    \caption{
    Contour map of the velocity-integrated $^{12}$CO (1--0) intensity image
    overlaid on the pseudocolor scale representations of 
    (a) the H$\alpha$ image,
    (b) the 8$\mu$m image,
    (c) the 24$\mu$m image,
    and 
    (c) the \ion{H}{1} image, respectively.
    Contour levels are 1, 3, 5, 7, and 9 times 1.0 Jy beam$^{-1}$ km s$^{-1}$. 
    Dashed circles indicate the approximate size of the NMA field of view ($\sim 60{''}$).  }
	\label {fig: a12_multic}
    %------------------
    % SCRIPT file is
    % @/home/hiko/work/paper7/scripts/batch_12co_multic
    %------------------
\end {figure*}
\ \ While $^{13}$CO peaks on the southern side are aligned with $^{12}$CO peaks, 
$^{13}$CO peaks on the northern sides are largely deviated from $^{12}$CO peaks.
In particular, the $^{12}$CO sources located around Point 3 (GMA center) 
lack clear counterpart in $^{13}$CO image.
As in the case of $^{12}$CO, the $^{13}$CO distribution on the northern side is spatially extended compared to the southern side. 
\subsection {Comparison with Multi-wavelength Data}
%\placefigure {fig: a12_multic}
%----------------------------------------------
% Multi-Color comparison
%----------------------------------------------
%\ \ Figure \ref {fig: } shows comparison with other wave length data. Comparison with H$\alpha$ image shows that while the southern emission is closely associated with HII regions, the northern emission is rather devoid of star forming regions. \\
%
%
%   \subsubsection{Ks-band}
%   \ \ Figure \ref{fig: 12co_k_a12} shows the 2MASS Ks image. 
%   The 45m $^{12}$CO (1--0) image and the NMA \& 45m combined $^{12}$CO (1--0) image are also shown with contours. The image is convolved with 10$^{''}$ beam to high-light the weak stellar arm. 
%   Since IC 342 is close to the Galactic plane (b = +10.579954$^{\circ}$), the NIR image is hampered by the foreground stars. Although large number of the foreground stellar images are removed from the Ks-band image, some of them cannot be removed because of image saturation. The yellow cross on the figure indicates the position of the saturated stellar image. \\
%   \ \ Comparison of the CO images with the Ks-band image shows that the GMA is located at the upstream side of the spiral arm. Meanwhile, the southern bright GMCs are located downstream of the GMA, close to the spiral arm. For clarity, the line of logarithmic spiral fitted on the $\log{R}-\theta$ plane in chapter 2 is drawn. The spiral phase  is determined as an angular offset $\phi$ from the logarithmic spiral. The lines of $\phi = 20^{\circ}, 10^{\circ}, -10^{\circ}$, and $-20^{\circ}$ are also drawn. 
%   %
\subsubsection{Star formation tracers}
%-----------------
% Ha, 8um, 24um
%-----------------
The $^{12}$CO image was compared with H$\alpha$, 8$\mu$m, and 24$\mu$m images to see the spatial relation between the distributions of molecular clouds and star forming regions.
Monochromatic H$\alpha$ image data were provided by \citet{Hernandez2005}.
Both the 8$\mu$m image taken with the Infrared Array Camera \citep[IRAC;][]{Fazio2004IRAC} and 
the 24$\mu$m image taken with the Multiband Imaging Photometer \citep[][]{Rieke2004MIPS} were 
retrieved from the ${\it Spitzer}$ archive. 
Resolution of the ${\it Spitzer}$ images is $\sim 1{''}.2$ for the $8\mu$m image and $\sim 6{''}$ for the $24\mu$m image.
Figure \ref{fig: a12_multic}(a)--(c) show the H$\alpha$ image, the 8$\mu$m image, and the 24$\mu$m image respectively.
All the images are overlaid with the $^{12}$CO (1--0) image.\\
%Overall distribution of star formation tracers resemble each other. 
\ \ It is apparent that while the clouds on the southern side are closely associated with 
their neighboring star forming regions
most of the clouds on the northern side seem to lack associated star forming regions.
In particular, around Point 3 (center of the GMA), 
little star formation activities are seen 
%only a few signs of star formation are 
seen both in the H$\alpha$ and the mid-infrared images. 
As we have seen before, the molecular gas distribution around the center of the GMA is 
much more smooth compared to in the southern side where molecular clouds are concentrated to form 
the narrow ridge.
To see whether such difference of the distribution and star formation activity in both sides of the observed region is related to the properties of clouds, cloud properties will be examined in the latter (Section \ref{sec: change_prop}).\\
%--------------
% 8um-PAH
%--------------
    %   \ \ Figure \ref{fig: a12ncomb_r5_m0_irac} shows comparison with IRAC 8$\mu{m}$ image. 
    %   The 8$mu{m}$ flux from HII region is dominated by polycyclic aromatic hydrocarbon (PAH) emission which is mainly excited 
    %   by UV photons and suggested to be attractive tracer of star formation. 
    %   The distribution of the 8$\mu{m}$ image resemble with the H$\alpha$ image. 
%--------------------------
% Buried star formation?
%--------------------------
%   Interesting feature is seen at point 1. 
%   While H$\alpha$ emission is not clear at the point, 8$\mu$m and 24$\mu$m image shows existence of discrete source. 
%   %   Around the point 3, no or little 8$\mu$m emission is seen. 
%   So, the sparseness of HII region in the GMA is not due to extinction. 
%   MIPS 24$\mu{m}$ image which traces the hot dust emission heated by HII region also shows similar trend (Figure \ref{fig: a12ncomb_r5_m0_mips}). \\
%
%-------------------------------------------------
% CO distirbutions and star formation activities
%-------------------------------------------------
%
\ \ Previous CO observations made with coarse spatial resolutions (300--1000 pc) 
often found the well-ordered spatial offsets between the molecular spiral arm and star-forming regions in 
other spiral galaxies \citep[e.g.,][]{VogelKulkarniScoville1988, RandLordHidgon1999M83}.
Also in \objectname{IC 342}, with the spatial resolution of $\sim 320$pc, the GMA is seen offset from 
the star-forming regions \citep[][]{Hirota2010IC34213CO}.
While on the other hand, seen with the 50pc resolution here, 
the separations between the clouds and the associated star-forming regions are small (mostly below the beam size).
Recent observations of the grand design spiral galaxy M51 also indicate similar result 
\citep[][]{Egusa2011M51}.
It is suggested that large GMA found with the previous coarse resolutions is 
a mixture of the both kinds of clouds,
which are clouds associated with and not with star forming region, 
as in \objectname{IC 342}.
%
%
%\subsubsection {H{\tiny {\rm I}}}
%\subsubsection {\mbox{\ion{H}{1}}}
\subsubsection {H{\scriptsize {\rm I}}}
\label{sec: 342HI}
%-----------------------------------
% scaling is 16.1722
%-----------------------------------
%--------------------
% HI
%--------------------
\ \ Figure \ref{fig: a12_multic} d) shows the comparison between the $^{12}$CO image and the \ion{H}{1} image. 
\ion{H}{1} data were retrieved from the Very Large Array archive and reduced with the Astronomical Image Processing System. 
The resultant \ion{H}{1} data cube had a spatial resolution of $23{''}.2 \times 20{''}.1$ and typical rms noise of $\sim$ 0.7 $\mathrm{mJy}$ beam$^{-1}$ for each channel. 
The surface mass density of \ion{H}{1} is calculated under the assumption of 
optically thin \ion{H}{1} emission.
Typical surface density of \ion{H}{1} over the observed field is $\sim 2--4$ $M_{\odot}$ pc$^{-2}$ and is 
smaller than that of molecular gas by an order of magnitude.
Though the spatial resolution of the \ion{H}{1} image is $\sim 20{''}$ and worse than the NMA image, 
it is apparent that there exists a hole at the center of the GMA. 
Around the \ion{H}{1} hole, distribution of \ion{H}{1} roughly coincides with that of \ion{H}{2} regions.
The coincidence of \ion{H}{1} with \ion{H}{2} regions indicates that the \ion{H}{1} clouds around the GMA might produced by the dissociation of H$_2$ by the UV radiation.
\subsection {Cloud identification and basic cloud properties}
\begin {figure} [htbp]
  \begin {center}
    %\plotone{img/a12ncomb_r5_m0_cl_xy.ps}
    \plotone{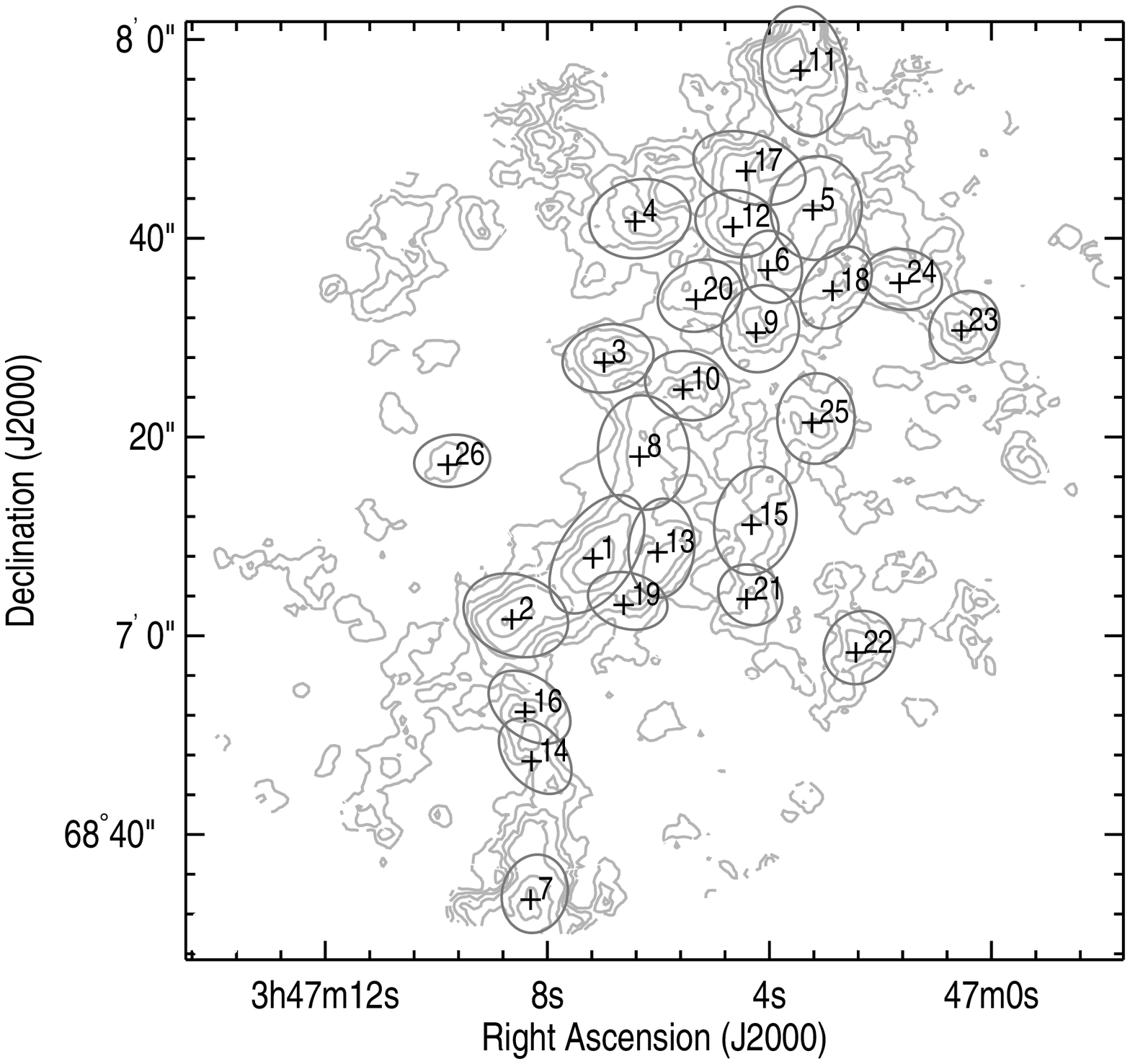}
	\caption {Locations of the identified GMCs overlaid on the velocity integrated $^{12}$CO (1--0) image. 
    Contour levels are the same as in Figure \ref{fig: m0_nma_combined}.
    % tclpvfile -> d(6,*) - rarr2 (diameter)
    Positions of the identified clouds are indicated with crosses.
    Ellipses indicate the FWHM sizes of the intensity distribution for each cloud.
    The FWHM sizes and position angle of the ellipses were determined by calculating the eigenvectors of 
    the intensity-weighted covariance matrix for each cloud 
    \citep[see][]{Koda2006Elongation, Rosolowsky2006BiasFree}.
    Clouds IDs are also indicated on the right of the crosses.  }
	\label {fig: a12ncomb_r5_m0_cl_xy}
  \end {center}
% /home/hiko/temp/temp.pro
\end {figure}
To investigate the properties of molecular clouds across the spiral arm, 
decomposition of individual cloud emission from the combined $^{12}$CO (1--0) data was attempted.
The CLUMPFIND algorithm \citep{Williams1994CLFIND} with some fine-tuning of parameters, which were proposed by \citet{RosolowskyBlitz2005}, was utilized.
Original code of the CLUMPFIND algorithm was targeted at identifying Galactic molecular cloud data observed with single-dish telescopes, in which beam size of the observation is comparable to the pixel size and S/N greatly differs from typical extragalactic observations.
%resolution unit size to pixel size ratio and signal to noise ratio differs from typical extragalactic cloud observations.
\citet{RosolowskyBlitz2005} performed cloud decomposition from the interferometric data of \objectname{M64}, which is a molecular rich galaxy with a similar distance to \objectname{IC 342}.
They proposed some modifications to be made to the original CLUMPFIND algorithm in applying the 
algorithm to extragalactic observation data. 
As the spatial resolution and the S/N of our data resembles that of \citet{RosolowskyBlitz2005}, 
we adopted some of the modifications proposed by the author.
\\
\begin {figure*} [htbp]
  \begin {center}
    %\plotone{img/a12ncomb_r5_m0_cl_xy.ps}
    %\plotone{img//342arm.ch.cl.p1.eps}
    \plotone{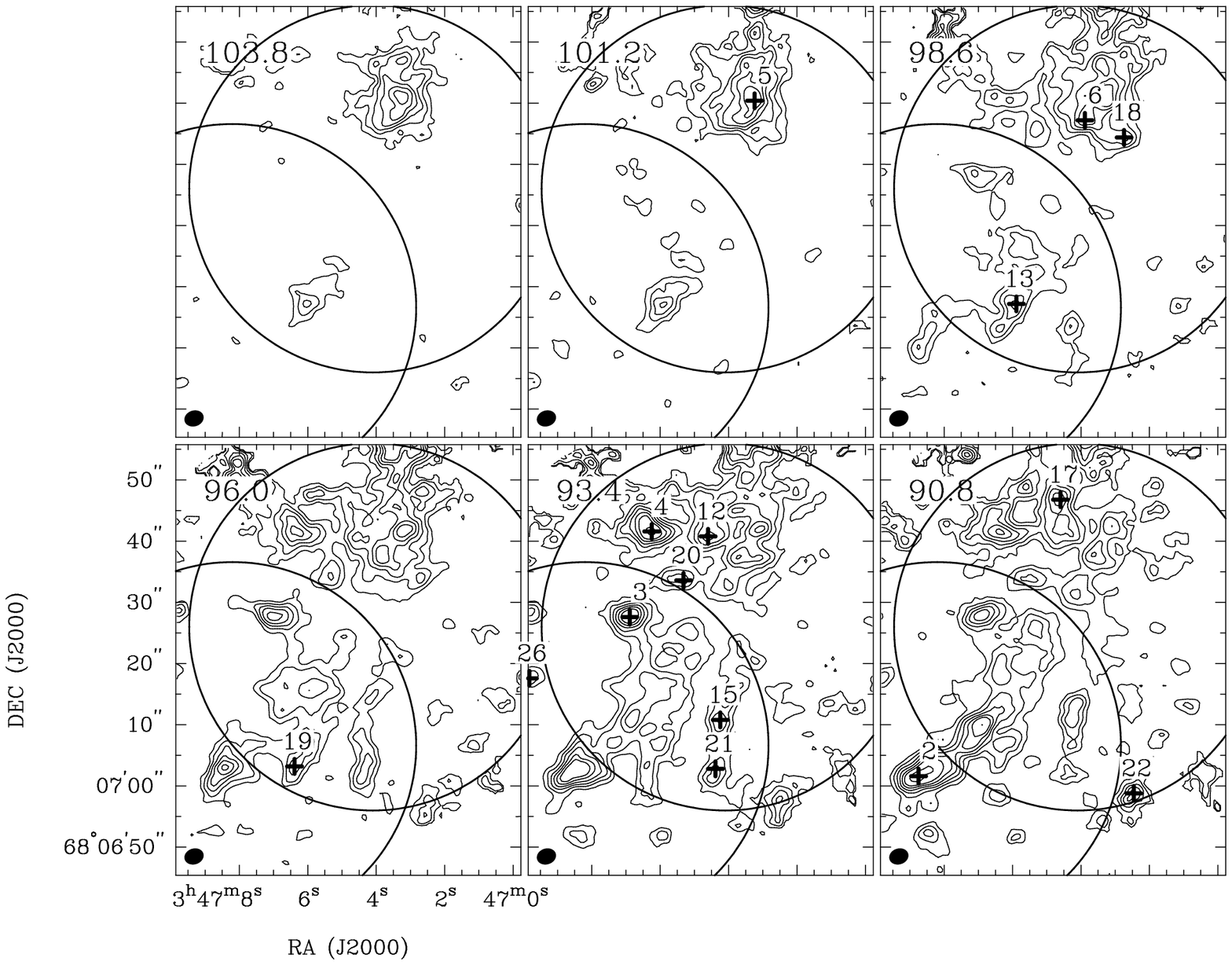}
	\caption {Channel maps of $^{12}$CO(1--0) line emission for the selected spatial and 
    velocity range of the data.  Peak positions and IDs of the identified clouds are indicated with crosses and associated labels.  Contour levels are 4, 6, 8, 10, 12, 14, 18, and 22 times 23 $\mathrm{mJy}$ beam$^{-1}$, 
respectively.  Label on the top left corner of each panel denotes the corresponding center velocity
($V_{\rm{LSR}}$ in km s$^{-1}$) for each channel.}
	\label {fig: a12ncomb_chmap_cl_xy}
  \end {center}
% /home/hiko/temp/temp.pro
\end {figure*}
\addtocounter{figure}{-1}
\begin {figure*} [htbp]
  \begin {center}
    %\plotone{img/a12ncomb_r5_m0_cl_xy.ps}
    \plotone{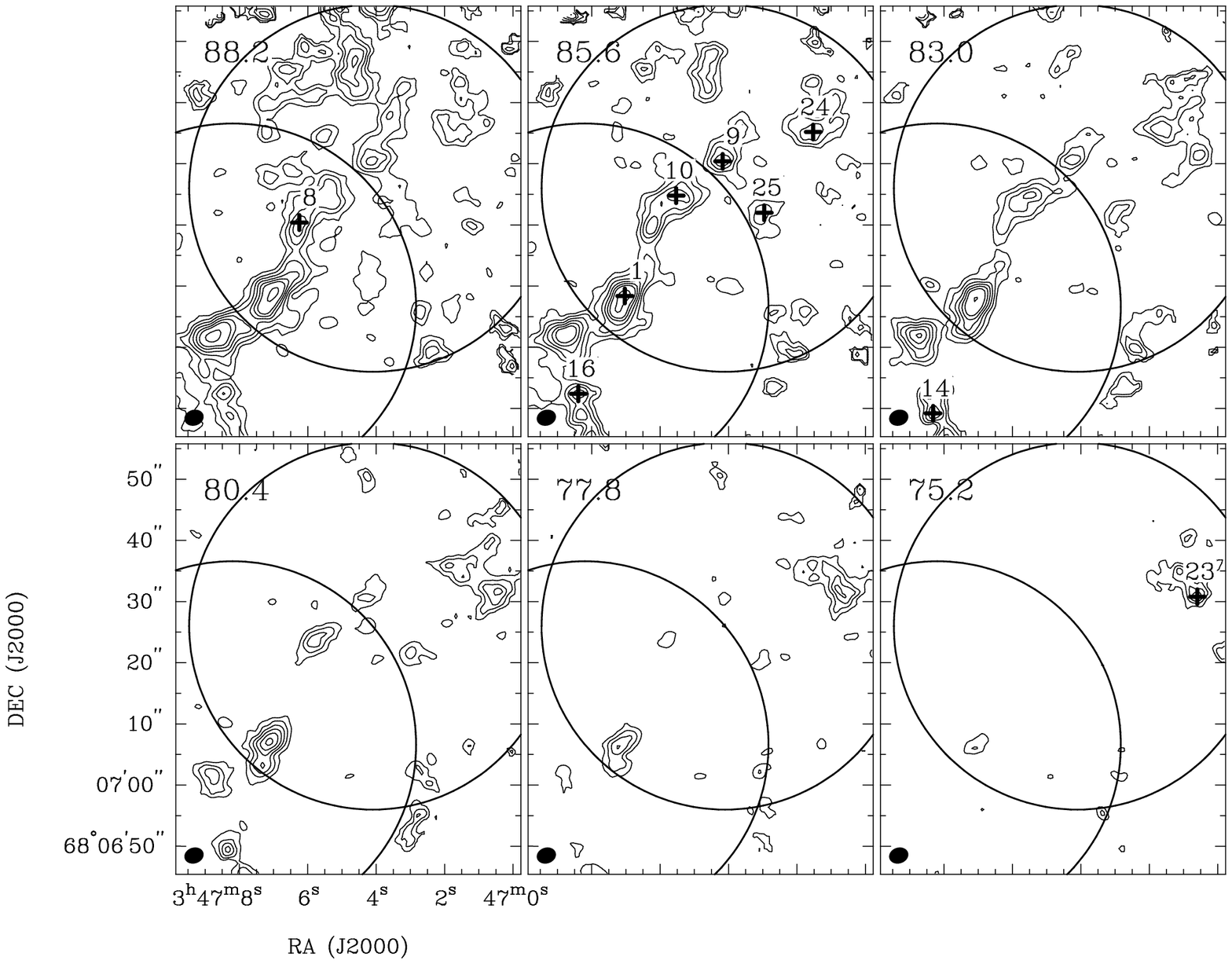}
	\caption {(Continued)}
	\label {fig: a12ncomb_chmap_cl_xy}
  \end {center}
% /home/hiko/temp/temp.pro
\end {figure*}
\ \ The modifications proposed by \citet{RosolowskyBlitz2005} consists of two points: 
one is to alter the manner of locating significant cloud peaks and another is to change distance metric along the velocity axis to account for the oversampling in spatial direction.
Among these two modifications, we had adopted only the latter one.
In the following, we will briefly mention about the modification we had adopted and not adopted. 
%We had applied both of the algorithm, original and modified versions of the CLUMPFIND, to our data.
First, we had adopted the change of distance metric.
As the spatial resolution of the combined data is $\sim$ 3${''}$ and the cell size of the data is 0.4${''}$, the data are heavily oversampled in the spatial direction compared to the velocity direction, 
in which the resolution is $\sim$ 5.2 km s$^{-1}$ and the pixel size is $\sim$ 2.6 km s$^{-1}$.  
To account for such oversampling, the distance metric 
used in the partitioning cloud boundary was modified \citep[see Appendix A of][]{RosolowskyBlitz2005}.
Second, we tested the modified procedure of finding significant local maxima.
While the original CLUMPFIND code searches and partitions data cube 
into each clump by a fixed intensity interval (usually taken as 2$\sigma$), 
the method proposed by \citet{RosolowskyBlitz2005} searches and extends clumps by every 0.5$\sigma$ 
but adopt 3$\sigma$ threshold in discarding 'false' local maxima. 
We tested the both algorithms and found that the results of the cloud participating are similar to each other, except for clouds around Point 3 (the GMA center).
While the original CLUMPFIND code produced three clouds around the point, the modified CLUMPFIND algorithm 
bundled the three clouds into a single large cloud complex. 
However, close examination of the results indicated that the three clouds bunched by 
the modified CLUMPFIND algorithm seem to be individual entity because of the different 
center velocity. 
Because of this, we did not adopt the modified procedure for finding local maxima in participating the data.\\
%So, we adopted the results produced by the original CLUMPFIND code.\\
%   Because of this, we did not adopt a modified procedure of finding local maxima
%   in participating data.\\
\ \ Identification of clouds from the combined NMA+45m data was performed with the following steps.
A data cube containing S/N value as a pixel value was generated by dividing the data cube before primary beam correction with rms noise level. 
The CLUMPFIND algorithm \citep{Williams1994CLFIND} with minor modifications mentioned above was applied to the S/N data cube with a 2$\sigma$ increment level and 2$\sigma$ lowest cutoff level. 
After partitioning the data cube into each clump, 
clumps located completely outside of the primary beam and clumps with peak temperature below $7\sigma$ were excluded from the analysis. 
The $7\sigma$ cutoff was set by the fact that over this level, 
both the algorithms tried here produced similar results, as have noted in the previous.
Finally, 26 GMCs were identified from the $^{12}$CO data cube.
Figure \ref{fig: a12ncomb_r5_m0_cl_xy} shows locations and approximate extent of the identified clumps 
projected on spatial directions. Figure \ref{fig: a12ncomb_chmap_cl_xy} shows channel maps of the selected regions from the data cube with peak positions of the identified clumps overlaid, to 
show how the identification of the clumps was executed.  \\ 
%
%Original CLUMPFIND code identify clumps from the data in the following way. First contour the data by fixed intensity step (usually taken as 2$\sigma$), and starting from the maximum contour level, mark the local peaks found over the contour level and define those as clump peaks, 
%Next, going down to the lower contour level step by step, to find new local peaks emerged and 
%extend the boundaries of existing clumps. When several clumps share the same pixels at some contour level, the contour level is defined as saddle point and 
%The clumps outside the primary beam were also removed. \\
%   \subsubsection {Basic Cloud Properties}
\ \ Properties of the identified clouds, namely, 
mean position, 
mean velocity, 
cloud radius ($R$), 
FWHM velocity width ($\Delta{V}$), 
and CO luminosity ($L(^{12}\rm{CO})$) 
were measured with the data cube corrected for primary beam attenuation.
The mean position and the mean velocity of the clouds were taken as the first moment of the emission. 
%
% Reff(w) = sqrt(sx(w) * sy(w)) / sqrt(8.0 * alog(2.0)) * 3.4 / sqrt(!pi)  
%
The effective radius of the identified clouds was calculated by 
%   \begin {equation}
%      Re = \frac{3.4}{\pi} 
%           \sqrt{ f(P) * \sigma_{x, obs}^2 - \sigma_{beam, maj}^2 } *
%           \sqrt{ f(P) * \sigma_{y, obs}^2 - \sigma_{beam, min}^2 }       
%   \end {equation}
\begin {equation}
    R= \frac{3.4}{\sqrt{\pi}}
            %\sqrt[4]{ f(P) * \sigma_{x, obs}^2 - \sigma_{beam, maj}^2 }
            %\sqrt[4]{ f(P) * \sigma_{y, obs}^2 - \sigma_{beam, min}^2 }       
            \left( \sigma_{x}^2 - \sigma_{\rm{beam, maj}}^2 \right)^{1/4}
            \left( \sigma_{y}^2 - \sigma_{\rm{beam, min}}^2 \right)^{1/4}, 
\end {equation}
where 
$3.4 / \sqrt\pi$ is a factor for converting the rms size into the radius of a spherical cloud 
\citep{Solomon1987Larson},  
$\sigma_{x}$, $\sigma_{y}$ is the second moment of the intensity distribution in spatial directions,
and $\sigma_{\rm{beam, maj}}$ and $\sigma_{\rm{beam, min}}$ is the rms size of the observed beam in major and minor axis directions, respectively.
As the rms sizes ($\sigma_x$, $\sigma_y$) are underestimated because of the 2$\sigma$ clipping level for each cloud boundary, correction should be made.
Often used method in this case is a Gaussian correction which assumes a Gaussian profile of the cloud emission and 
boosts the measured properties with the factor determined by the ratio of the peak temperature to the truncation level 
\citep[][]{Oka2001GCMC, Bolatto2003SMC, RosolowskyBlitz2005, Rosolowsky2007M31}. 
The correction was made following the analytic expression given by \citet{RosolowskyBlitz2005}, 
which boosts the rms size by factor of at most 1.2 for our data. 
Line width of the clouds was taken as the FWHM size of the cloud profile, also corrected with the 
Gaussian correction, and corrected for resolution bias by
subtracting the velocity resolution ($\sim$ 2.6 km s$^{-1}$) in quadrature.
%where $D$ is the distance to \objectname{IC 342} \citep[$=3.3$Mpc,][]{SahaClaverHoessel2002}, 
%and $\Delta_x$, $\Delta_y$ is the FWHM cloud size in x and y directions, respectively.
%The FWHM cloud sizes ($\Delta_{x}, \Delta_{y}$) and velocity width ($\Delta_{v}$) were calculated from the second moment of the emission ($\sigma_x, \sigma_y, \sigma_v$)
%in each direction, e.g., 
%$\Delta_i = \sqrt{8\log{2}} \sigma_i$ ($i = x, y, v$).
%and $\sigma_x$, $\sigma_y$ the rms spatial dispersions. 
CO luminosity was taken as the summation of the emission within each cloud boundary corrected with the Gaussian correction.
Luminosity based cloud mass ($M_{\rm{CO}}$) was derived from the CO luminosity by applying the standard conversion factor noted in 
the previous section (Section \ref{sec: 12co_distribution}).
%\ \ The properties of identified clouds, namely, mean position, mean velocity, effective radius, FWHM velocity width $\Delta{V}$, and total CO luminosity were reduced. 
%\begin {equation}
%L_{\rm{CO}}=D^2 \sum_{x}\sum_{y}\sum_{v}{T}
%\end {equation}
%
The virial mass of a spherical cloud with density profile $\rho \propto{r}^{-n}$ is written as \citep[e.g.,][]{MacLaren1988VirialMass}
\begin {equation}
\frac{M_{\rm{vir}}}{M_\odot}=126 \frac{5-2n}{3-n}\left(\frac{R}{\rm{pc}}\right)
\left(\frac{\Delta{V}}{\rm{km\ s}^{-1}}\right)^2 .
\end {equation}
The density profile of $n = 1$ was assumed in calculating the virial mass.\\
\ \ Large scale kinematics of the galactic disk such as the galactic rotation and streaming motions 
may bias the measured line width and hence the virial mass. 
To check the possible influence of this, the velocity shear across each cloud was estimated by
\begin {equation}
\Delta{V}_{\rm{shear}}^2=
\frac{{\sum}I(x_i, y_i)
    \Bigl[
        v_{\rm{rot}}(x_i, y_i) - v_{\rm{rot}}(x_0, y_0)
    \Bigr]^2}
{{\sum}I(x_i, y_i)},
\end {equation}
where $x_i$, $y_i$ is the position of each pixel in the cloud, $v_{\rm{rot}}(x_i, y_i)$ is the circular rotational velocity at point ($x_i$, $y_i$), and $v_{\rm{rot}}(x_0, y_0)$ is the circular rotational velocity at the mean position of the cloud ($x_0$, $y_0$). 
An estimated amount of $\Delta{V}_{\rm{shear}}$ was smaller than the measured line widths by nearly an order of magnitude and turns out to be negligible for our measurement. \\
\ \ Table \ref{tab: ic342_GMC_props} denotes the basic properties of the identified molecular clouds. 
The summation of the mass of the identified clouds is $\sim$ 4.4 $\times 10^7 M_{\odot}$, 
which comprises $\sim$ 60$\%$ of the total molecular mass within the observed region. 
The relations between the basic properties of the identified clouds are indicated in Figure
\ref{fig: cl_prop1_a12}(a)-(d). As the Gaussian correction is the largest source of uncertainty for the measurement of cloud properties, uncorrected values are also indicated for comparison. 
\begin {deluxetable*}{lccccccccc}
    \tablecaption {Basic Molecular Cloud Properties}
    \tablewidth{0pt}
    \tablecolumns{10}
    \tabletypesize{\small}
    \tablehead{
        % First Row
        \colhead{ID} &
        %\multicolumn{2}{c}{position\footnotemark[1]} & 
        \multicolumn{2}{c}{Position\tablenotemark{a}} & 
        \colhead{$V_{\rm{LSR}}$} & 
        \colhead{$T_{\rm{peak}}$} & 
        \colhead{$R$} &
        \colhead{$\Delta{V}$} & 
        \colhead{$M_{\rm{CO}}$} & 
        \colhead{$M_{\rm{vir}}$} \\
%        \colhead{$\nu{L}_{\nu}(8\mu{m})$\tablenotemark{b}} \\
        % Second Row
        \colhead{}  &
        \multicolumn{2}{r}{(arcsec, arcsec)}         & 
        \colhead{(km s$^{-1}$)} & 
        \colhead{(Jy beam$^{-1}$)} &
        \colhead{(pc)} &
        \colhead{(km s$^{-1}$)} &
        \colhead{($10^6$ $M_\odot$)} &
        \colhead{($10^6$ $M_\odot$)} 
%        \colhead{($10^5 L_{\odot}$)}
    }
%\tablehead{
%    \colhead{} &
%}
    \startdata
   1 &   16.8 &  --17.7 &   87.2 &  0.47 &   62.0 &   12.4 &    3.58 &    1.81 \\
   2 &   25.0 &  --23.8 &   89.5 &  0.47 &   57.2 &   11.6 &    3.32 &    1.47 \\
   3 &   15.7 &    1.8 &   92.6 &  0.40 &   44.3 &    8.9 &    1.32 &    0.66 \\
   4 &   12.6 &   15.9 &   93.1 &  0.44 &   54.5 &    9.2 &    1.84 &    0.87 \\
   5 &   --5.3 &   17.0 &  100.5 &  0.41 &   62.0 &   14.1 &    3.66 &    2.34 \\
   6 &   --0.8 &   11.0 &   97.0 &  0.32 &   32.4 &   14.6 &    1.37 &    1.31 \\
   7 &   23.1 &  --51.8 &   79.4 &  0.62 &   37.2 &    9.4 &    2.42 &    0.63 \\
   8 &   12.1 &   --7.6 &   90.9 &  0.28 &   69.5 &   12.1 &    2.39 &    1.91 \\
   9 &    0.4 &    4.8 &   87.1 &  0.27 &   51.5 &   13.0 &    1.46 &    1.66 \\
  10 &    7.7 &   --0.9 &   87.2 &  0.27 &   45.6 &   10.9 &    1.46 &    1.03 \\
  11 &   --4.1 &   30.9 &   96.3 &  0.60 &   66.5 &   10.2 &    3.98 &    1.31 \\
  12 &    2.7 &   15.3 &   95.4 &  0.29 &   42.6 &   15.2 &    1.80 &    1.87 \\
  13 &   10.3 &  --17.1 &   96.6 &  0.24 &   52.0 &   16.0 &    1.82 &    2.52 \\
  14 &   23.0 &  --38.0 &   84.1 &  0.29 &   38.9 &    9.8 &    0.83 &    0.71 \\
  15 &    0.9 &  --14.4 &   92.1 &  0.24 &   65.3 &   11.1 &    1.34 &    1.53 \\
  16 &   23.6 &  --33.1 &   87.9 &  0.26 &   43.5 &   14.8 &    0.87 &    1.80 \\
  17 &    1.4 &   20.9 &   94.3 &  0.31 &   55.9 &   12.8 &    2.11 &    1.74 \\
  18 &   --7.3 &    8.9 &   97.3 &  0.24 &   44.9 &   13.4 &    1.04 &    1.52 \\
  19 &   13.7 &  --22.4 &   93.2 &  0.22 &   39.9 &   16.2 &    0.99 &    1.98 \\
  20 &    6.4 &    8.1 &   94.1 &  0.20 &   52.7 &   12.3 &    1.10 &    1.50 \\
  21 &    1.3 &  --21.8 &   93.4 &  0.21 &   35.1 &    9.1 &    0.57 &    0.55 \\
  22 &   --9.7 &  --27.1 &   90.8 &  0.29 &   40.7 &   10.8 &    0.92 &    0.89 \\
  23 &  --20.3 &    5.0 &   77.2 &  0.26 &   41.5 &   13.7 &    1.02 &    1.48 \\
  24 &  --14.1 &    9.7 &   86.2 &  0.23 &   42.2 &    9.8 &    0.94 &    0.77 \\
  25 &   --5.2 &   --4.2 &   88.5 &  0.18 &   62.2 &   14.9 &    1.33 &    2.62 \\
  26 &   31.4 &   --8.4 &   92.9 &  0.18 &   38.1 &    9.6 &    0.38 &    0.67 \\
    \enddata
    \tablenotetext{a}{Offset from $3^{\rm{h}}$$47^{\rm{m}}$$4^{\rm{s}}.1$, $68^{\circ}$$7{'}$$26{''}$} \\
    %\tablenotetext{b}{Total 8$\mu$m flux within each cloud boundary (see \ref{sec: star_formation_rate})}
    \label{tab: ic342_GMC_props}
\end {deluxetable*}
%
%
%
%\placefigure {fig: a12ncomb_r5_m0_cl_xy}
%\placefigure {fig: cl_prop1_a12}
%-------------------------------------
% Cloud-properties
%-------------------------------------
%
\begin {figure} [htbp]
  \begin {center}
	%\plotone{img/cl_prop1.eps}
	\plotone{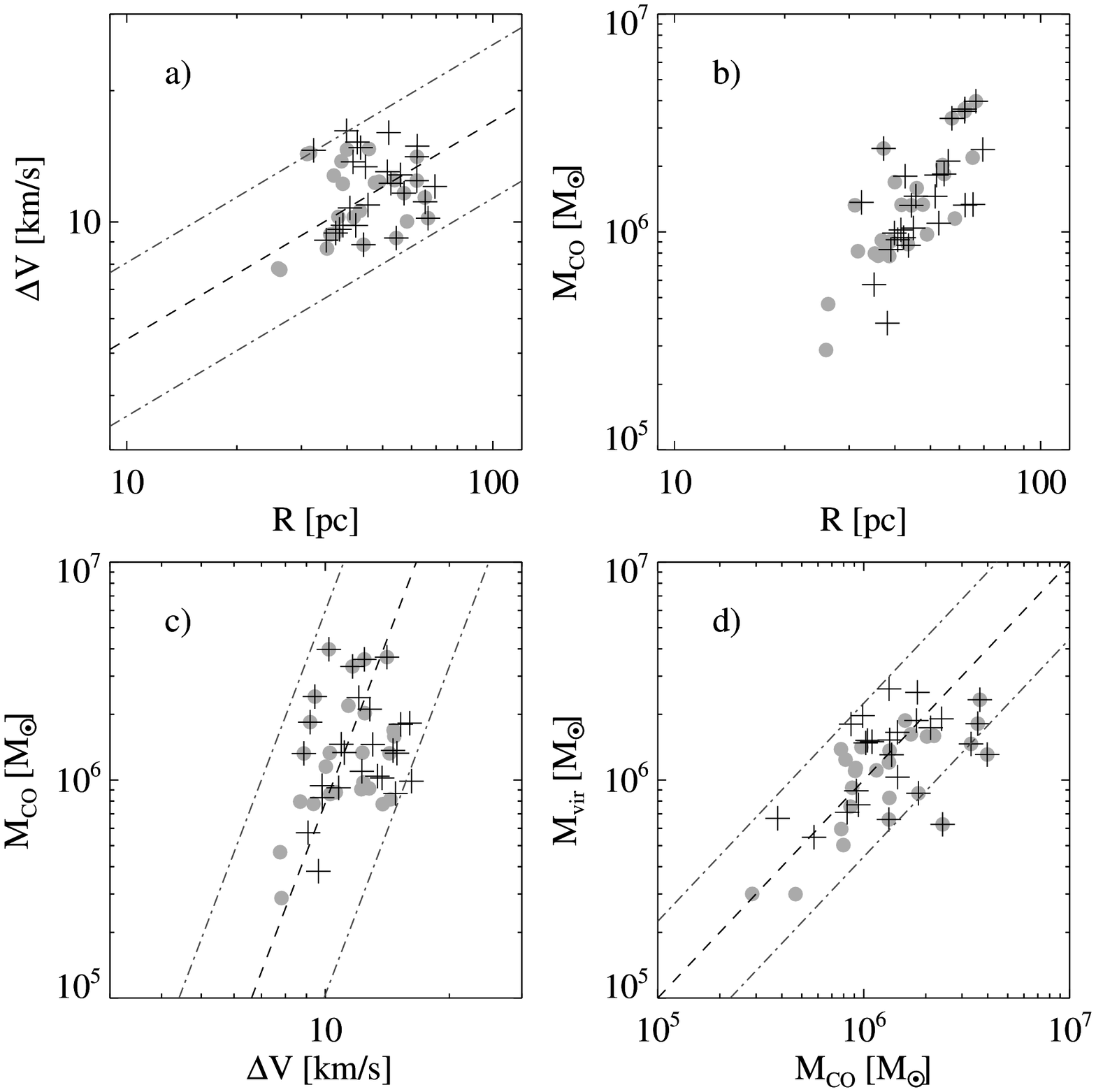}
	\caption {
    (a) $\Delta{V} - R$ plot for the identified GMCs. 
    Crosses indicate the values corrected with the Gaussian correction, and 
    gray filled circles indicate the uncorrected values. 
    Broken line indicates the scaling relation of \citet{Solomon1987Larson}.
    For comparison, the scaling relations with different coefficients (multiplied by 1.5 and 1/1.5, respectively) are 
    indicated with dash-dotted lines.
    (b) Same as (a) but for $M_{\rm{CO}} - R$. 
    (c) Same as (a) but for $M_{\rm{CO}}- \Delta{V}$.
    (d) Same as (a) but for $M_{\rm{vir}} - M_{\rm{CO}}$. }
	\label {fig: cl_prop1_a12}
  \end {center}
% see memo_342arm07_robust0.txt
\end {figure}
\subsection {Spatial Relation Between Star Formation Activity and Spiral Arm}
\begin {figure} [htbp]
	%\plotone {img/cl_8mu_flux_histo.eps}
	\plotone {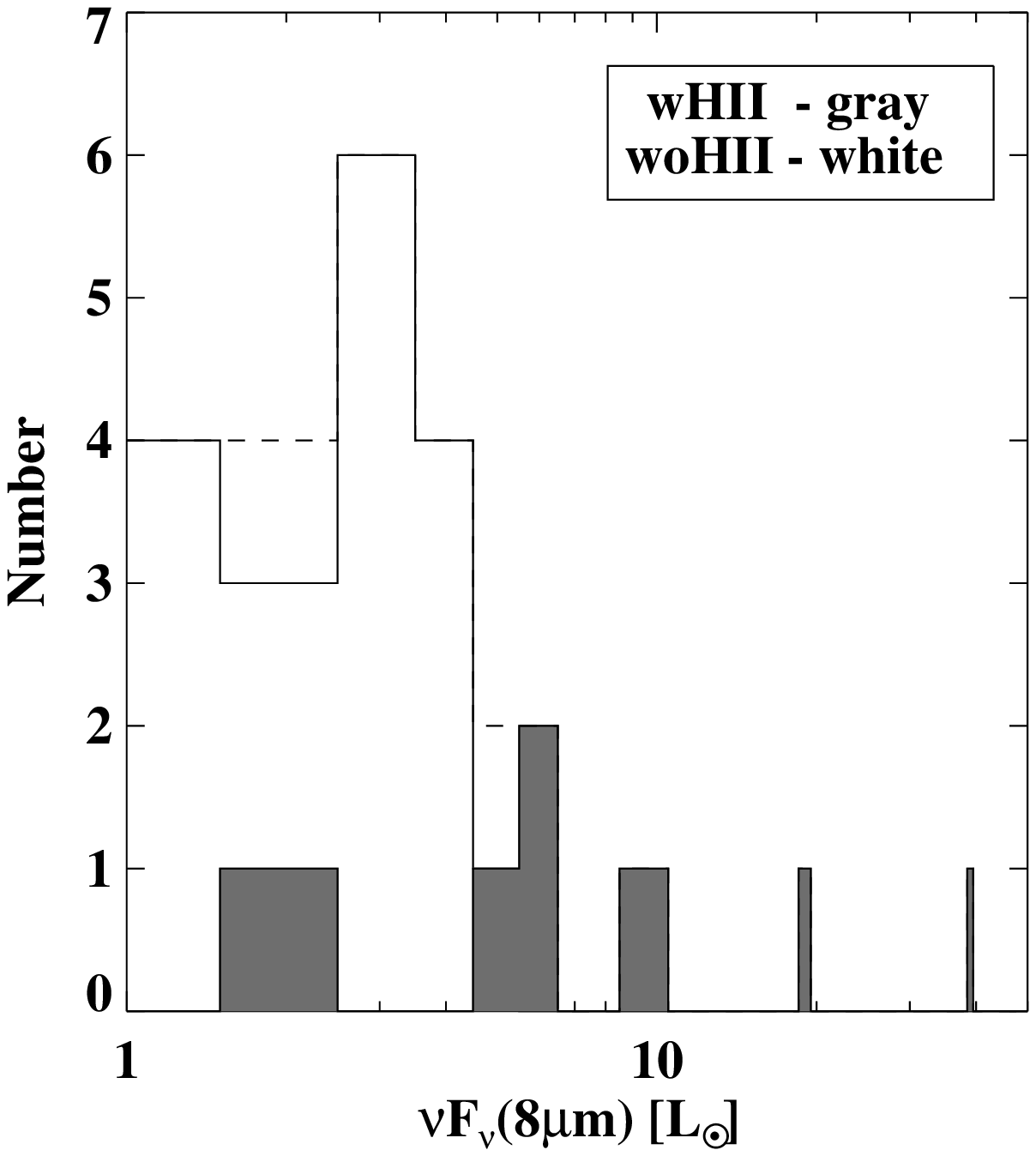}
	%\includegraphics [scale=0.9, trim=70 0 0 0 ] {img/a12_alpha_flux.ps}
	%\caption {The histogram of the basic properties of the GMCs, namely, radius, velocity dispersion, and luminosity mass. Red, and blue solid lines indicate the histogram of the "wHII" GMCs and "woHII" GMCs, respectively. Red and blue broken lines indicate the mean value of each category. }
	\caption {
    Histogram of the 8$\mu$m flux measured within each cloud boundary 
    of the "wHII" GMCs (gray area)
    and the "woHII" GMCs (white area).}
	\label {fig: cl_8mu_flux_histo}
% /home/hiko/temp/temp.pro
\end {figure}

\begin {figure} [htbp]
  \begin {center}
	%\plotone {img/a12ncomb_m0_cl_HII.eps}
	\plotone {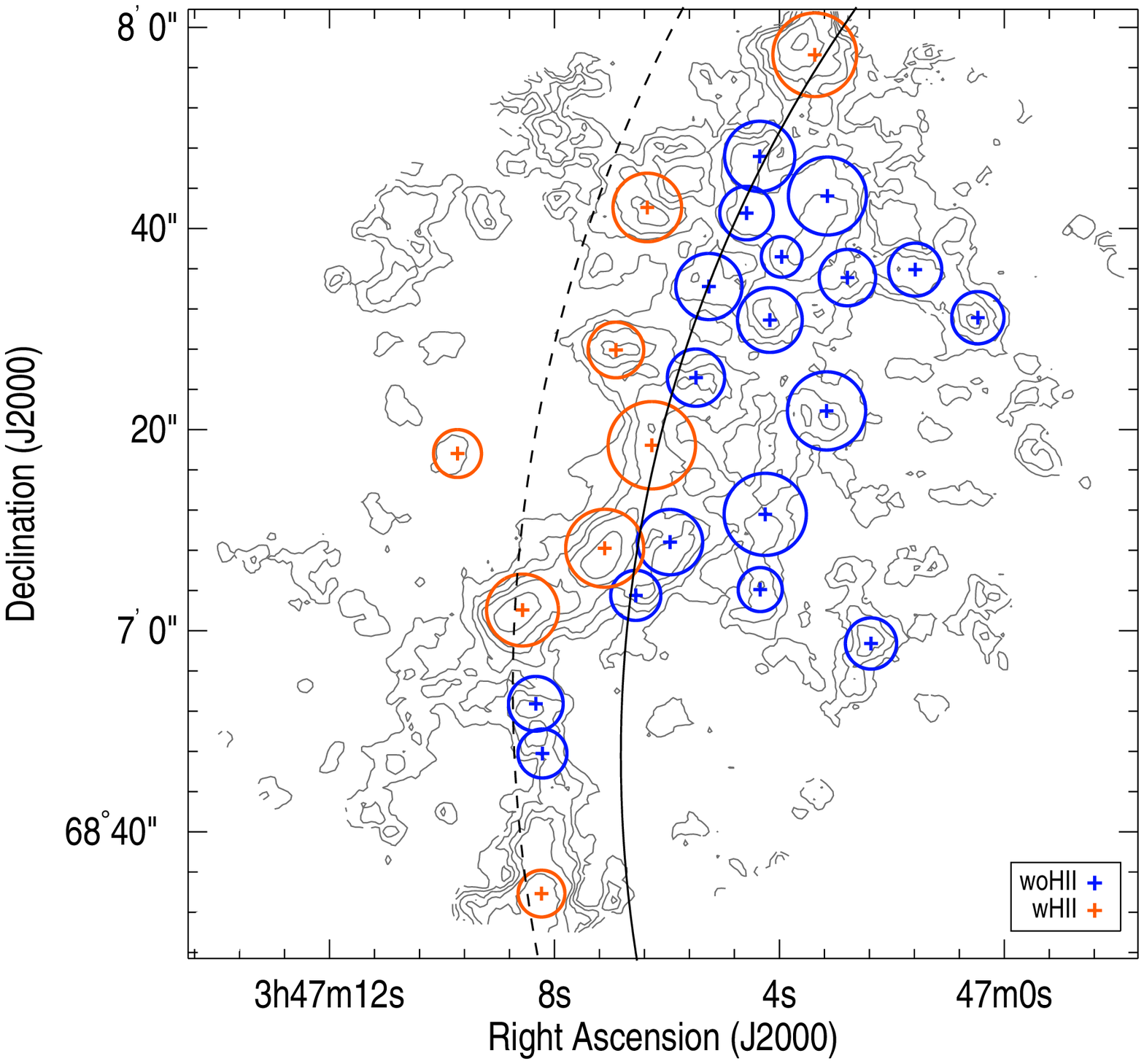}
	\caption {
    Locations of the identified GMCs superposed on contour map of the velocity integrated $^{12}$CO (1--0) image.
    Contour levels of the $^{12}$CO images are same as in Figure \ref{fig: m0_nma_combined}.
Crosses and circles indicate the locations and the sizes of the GMCs. 
Radii of the circles are taken as an effective radius of the GMCs.
    Red markers represent the GMCs associated with \ion{H}{2} region ("wHII" clouds), 
    while blue markers represent the GMCs without \ion{H}{2} region ("woHII" clouds).
    Lines of constant spiral phase are indicated with
    a dashed line ($\phi = 0^{\circ}$) and 
    a solid line ($\phi = -13^{\circ}$).
    }
	\label {fig: cl_guide_HII}
  \end {center}
% /home/hiko/temp/temp.pro
\end {figure}

%\subsection {Variation of Cloud Properties Across the Spiral Arm}
\label {sec: classification of gmcs}
\ \ The comparison between the $^{12}$CO image and the star formation tracers indicated that 
while some of the clouds seem to lack associated star forming activity,
the rest of the clouds are closely associated with accompanying star forming regions.
To check whether the properties of the clouds change with associated star forming activity, 
the identified GMCs were divided into two groups according to their star formation activity.\\
\ \ The H$\alpha$ and the mid-infrared images were examined for the classification.
Although the 24$\mu$m band image is preferable in tracing the star formation rate 
\citep{Calzetti2005ApjM51, Calzetti2007}, 
coarse resolution of the 24$\mu$m image ($\sim$ 6${''}$) hampers 
identification of the correspondence between star-forming regions and the identified clouds.
Because of this, the star-forming regions were identified from the IRAC 8$\mu$m image because of its higher resolution ($\sim$ 2${''}$) compared to the 24$\mu$m image.
There is two possible bias in tracing star formation with the 8$\mu$m image:
although the 8$\mu$m band is dominated by 
polycyclic aromatic hydrocarbon (PAH) emission mainly excited by photon-dominated region around star-forming regions, 
there is little but certain amount of stellar contribution to the 8$\mu$m band 
\citep[e.g.,][]{Helou2004NGC300} and 
the scaling between the star formation rate and PAH emission is slightly deviated from linear
possibly because of the
presence of  diffuse PAH emission excited by ambient radiation field 
\citep[e.g.,][]{Calzetti2005ApjM51}.
However, as we are interested in locating star-forming regions and not in 
measuring the exact star formation rate, this point is not a severe deficit.\\
    %   \ \ Figure \ref{fig: a12ncomb_r5_m0_irac} shows comparison with IRAC 8$\mu{m}$ image. 
    %   The 8$mu{m}$ flux from HII region is dominated by polycyclic aromatic hydrocarbon (PAH) emission which is mainly excited 
    %   by UV photons and suggested to be attractive tracer of star formation. 
    %   The 8$mu{m}$ flux from HII region is dominated by polycyclic aromatic hydrocarbon (PAH) emission which is mainly excited 
\ \ As intensity distribution of star-forming regions seems to have distinct outline, 
two-dimensional version of the CLUMPFIND algorithm \citep{Williams1994CLFIND} was
utilized in defying the boundary of each star-forming region.
Partitioning of the image was done by every 1.4 MJy str$^{-1}$ ($\sim 2\sigma$) step down to 7 MJy str$^{-1}$ level ($\sim 10\sigma$).
The lowest boundary level was determined to 
keep away from diffuse $8\mu$m emission which shall not be  
related with current massive star formation activity.  
The distribution of the identified star forming regions was also confirmed by comparing with the H$\alpha$ image.\\
\ \ The GMCs were divided into two group according to whether they are associated with the identified star forming regions or not.
If a GMC overlaps with the identified star forming region and 
separation between the center of the GMC and that of the associating star forming region is within the beam size, 
the GMC is categorized as being "associated with star formation". 
Those "GMCs with \ion{H}{2} region" are hereafter referred to as
wHII" GMCs.
On the other hand, the rest of the GMCs were categorized as "GMCs without \ion{H}{2} region" and termed as "woHII" GMCs.
For most of the "wHII" GMCs, associated star forming regions are 
seen in both the 8$\mu$m and the H$\alpha$ images, 
except for a GMC located near Point 1.
At Point 1, there exists an accompanying bright 8$\mu$m source but 
there is no clear counterpart in the H$\alpha$ image.\\
  % %-------------------------------------
  % % scripts/batch_irac8um_clfin2d.pro
  % %-------------------------------------
\ \ To qualitatively ensure the classification, 8$\mu$m flux was measured within the boundary of each cloud.
Figure \ref{fig: cl_8mu_flux_histo} shows a histogram of the measured 8$\mu$m flux.
Although there is a "wHII" GMC which shows exceptionally low 8$\mu$m flux compared to other "wHII" GMCs (cloud-26), 
most of the "wHII" GMCs show higher 8$\mu$m flux compared to the "woHII" GMCs.
The low 8$\mu$m flux for the cloud-26 is due to the fact that 
the cloud is offset from the associated \ion{H}{2} region and 
the 8$\mu$m flux is measured only within the cloud boundary.
For the rest of the "wHII" clouds, offsets between each cloud and the associated star-forming region are not so large compared to the cloud-26.
The 8$\mu$m flux histogram seems to ensure the classification made here.
\\
\ \ Figure \ref{fig: cl_guide_HII} shows the spatial distribution of the classified GMCs. 
As in other grand design spiral galaxies, the spiral arm in \objectname{IC 342} has an exponential nature; 
the spiral arm extends almost linearly 
on the $\log{R}$-$\theta$ diagram
\citep{Sato2006, Hirota2010IC34213CO}.
Lines of the constant spiral phase ($\theta=0^{\circ}, -13^{\circ}$) are indicated in the figure for comparison.
The comparison of the distribution of the GMCs with respect to the spiral phases shows that 
the "wHII" GMCs are located downstream of the "woHII" GMCs. 
If the line of $\theta = -13^{\circ}$ is taken as the partition line, 
the GMCs upstream and downstream of the line show the difference 
in terms of the star formation activity. 
The difference in the star formation activity within each GMC seems to be 
well related with the position with respect to the spiral arm;
while all of the clouds upstream of the line are "woHII" GMCs,
most of the clouds downstream of the line are "wHII" GMCs. 
This is in accordance with the prediction of density wave induced/regulated star formation.
Hereafter, we will inspect the change of the molecular gas properties 
according to the classification defined.\\
%
%
%\placefigure {fig: cl_8mu_flux_histo}
%-------------------------------------
% 8um flux histogram
%-------------------------------------
%\placefigure {fig: cl_guide_HII}
%-------------------------------------
% Cloud props - wHII - woHII
%-------------------------------------
\subsection {Line ratio}
\begin {figure} [htbp]
  \begin {center}
%	\includegraphics [scale=0.9, trim=50 0 0 0 ]{/home/hiko/reduction/img07/ratio_1312_nma_12co.ps}
    %\plotone{img/ratio_1312_nma_c5_cl_overlaid.eps}
	\plotone {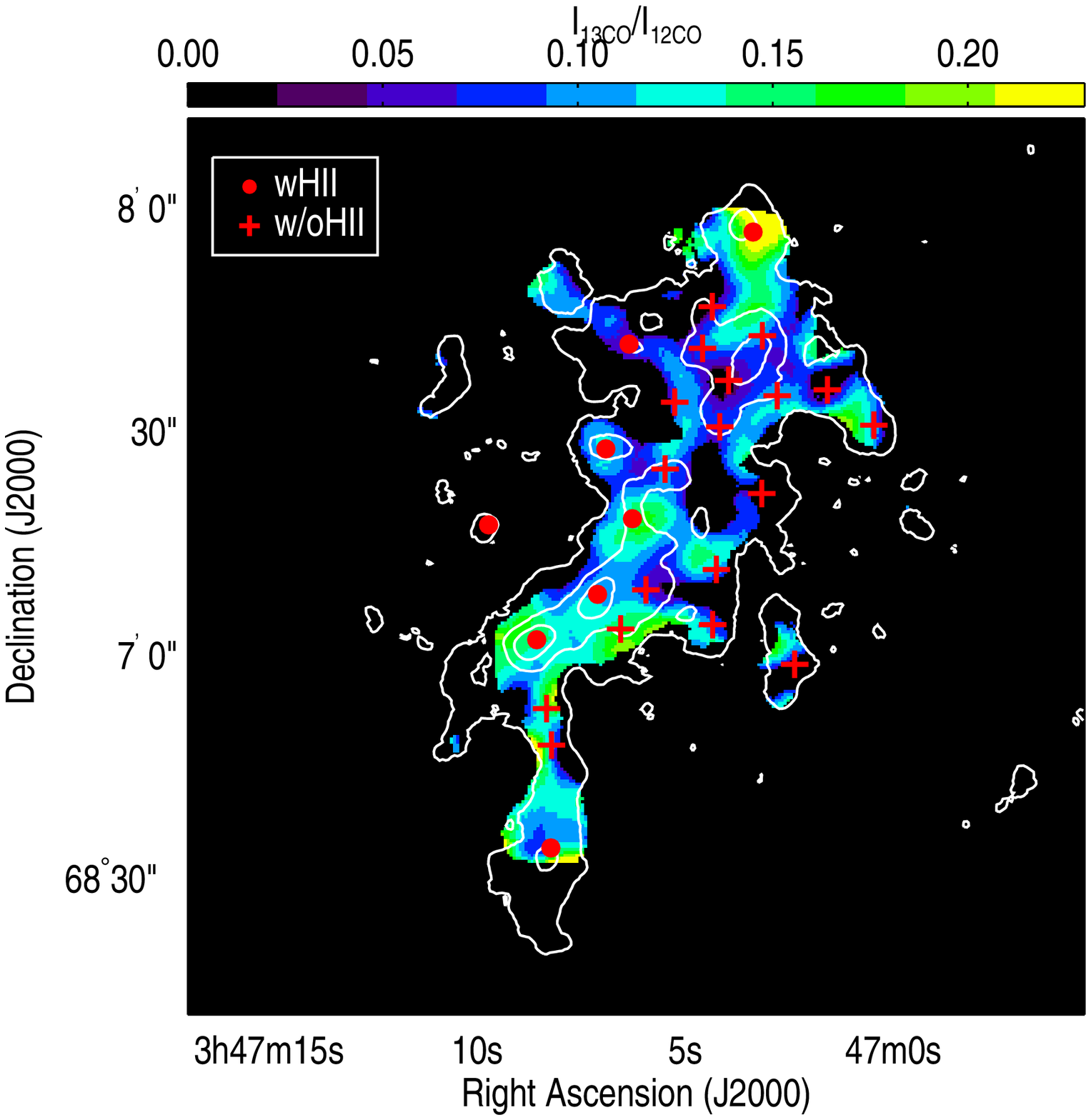}
  \caption {Color representation of $R_{13/12}$ superposed by contour map of $^{12}$CO (1--0) image.
  Contour levels are 1, 3, 5, and 7 times 1.5 Jy beam$^{-1}$ km s$^{-1}$. 
  Both images were made from the data cube smoothed to $5{''}$ resolution. 
  Locations of the "wHII" and the "woHII" GMCs are indicated with filled circles and crosses, respectively. Errors of the $R_{13/12}$ are typically $\sim$ 0.01 at the bright peaks and $\sim$ 0.02 at the rest regions. 
  %Red crosses indicated the local maxima of $^{12}$CO (1--0), which is marked in Figure \ref{fig: m0_nma_combined}. }
  }
     % scripts/batch_ratio_1312_nma_c5.pro 
	  \label {fig: ratio_1312_nma}
  \end {center}
\end {figure}

\ \ Because of the low critical densities of the $^{13}$CO (1--0) and the $^{12}$CO (1--0) lines 
($n_{\rm{H}_2} < 10^3$ cm$^{-3}$) and the large difference of the optical depth between both the lines, 
the line ratio between both the lines 
($I$($^{13}$CO (1--0))/$I$($^{12}$CO (1--0)) $\equiv R_{13/12}$) could be used as a probe which 
distinguishes discrete cloud structures such as ridge of GMCs from diffuse cloud envelopes.
It is widely known from the observations of Galactic molecular clouds that
while $R_{13/12}$ is high at the center of discrete GMCs \citep[1/3.7; ][]{Polk1988},
it is low (1/10--1/20) at the peripheral regions of GMCs \citep{Sakamoto1994Orion2to1} and in small translucent clouds or diffuse high-latitude clouds \citep{BlitzStark1986, MagnaniBlitz1985, KnappBowers1988}. \\
%
%
% GMAの中心は以上にR13/12が低い
%   => Cloud sizeとしてどれぐらい？
\ \ Previous extended mapping of \objectname{IC 342} in the $^{13}$CO (1--0) found 
that there exist variations of $R_{13/12}$ in the disk of \objectname{IC 342} \citep{Hirota2010IC34213CO}.
$R_{13/12}$ was found to be  low ($\sim$ 0.1) at the center of the GMA and both the bar ends compared to the other disk regions ($0.14$--$0.2$).
It was also found that the star formation activity in such low $R_{13/12}$ regions are low compared to the high $R_{13/12}$ regions. 
The spatial relation between $R_{13/12}$ and the star formation suggests that unlike in the starburst galactic centers \citep[e.g.,][]{Paglione2001}, the temperature variation is not a main cause of the $R_{13/12}$ variations in the disk region. 
Likely explanation of the $R_{13/12}$ variations in the disk is that 
$R_{13/12}$ reflects the fraction of the diffuse molecular component within the beam.
The diffuse molecular component refers to molecular cloud with 
low column and volume densities. \\
\ \ A map of $R_{13/12}$ was made with the following procedures.
First, to improve the S/N, both the $^{12}$CO and $^{13}$CO data cubes were smoothed to 5${''}$ resolution ($\sim 80$ pc). 
Next, masking data were made from the $^{12}$CO data cube following the same procedure described in Section \ref{sec: 13co_distribution} and were applied to the $^{13}$CO data cubes. 
Validity of the mask was checked with the same procedure performed in Section \ref{sec: 13co_distribution}.
Finally, the $R_{13/12}$ value for each point was calculated using the masked data.\\
\ \ 
\ \ Figure \ref{fig: ratio_1312_nma} shows the map of $R_{13/12}$ compared with the $^{12}$CO (1--0) map.
$R_{13/12}$ varies over the observed fields from $\sim 0.06$ to $\sim 0.25$.
An error of the ratio was estimated following the propagation of the error and 
was $\sim$ 0.01 at the brightest $^{12}$CO peaks (Point 1--3) and typically below 0.02 at the rest.
%--------------------
%   -------------------
%   Error estimate
%   -------------------
%   
% 342_13co/batch_make_ratio.c5.pro
%
$R_{13/12}$ is rather low on the northern side compared to the southern side. 
Especially, at the center of the GMA, $R_{13/12}$ marks low value ($\sim$ 0.06 $\pm$ 0.01), 
which is by far small compared to the ridge of GMCs \citep[1/3.7,][]{Polk1988} and 
close to the values found in diffuse clouds \citep{BlitzStark1986, MagnaniBlitz1985, KnappBowers1988}. 
On the other hand, southern clouds show higher $R_{13/12}$ ($\sim$ 0.16 $\pm$ 0.01), 
suggesting discrete nature of the clouds. \\
\ \ Comparing the distribution of the "wHII" and "woHII" clouds with the observed 
$R_{13/12}$ variation shows that there is a possible tendency that 
while $R_{13/12}$ is lower in the "woHII" GMCs, it is higher in the "wHII" GMCs. 
The configurations of the $R_{13/12}$ and star formation distributions 
exclude temperature variation as a primary factor for the $R_{13/12}$ variations.
The configurations also likely exclude abundance variation induced by 
selective photo dissociation of rarer isotopic species.
As the observed size scale of the $R_{13/12}$ variation is an order of 100pc, 
enrichment of $^{13}$C through stellar processing conflicts in terms of timescale.
Thus, the variation of the $^{13}$CO/$^{12}$CO abundance ratio is also unlikely.
As suggested by \citet{Hirota2010IC34213CO}, 
the most likely explanation for the $R_{13/12}$ variation in the disk of \objectname{IC 342}
is the variation of fraction of the diffuse molecular component. \\
\ \ Formation of dense cloud core ($n_{\rm{H}_2} \sim 10^6$ cm$^{-2}$) is required for the onset of star 
formation. 
Very low $R_{13/12}$ values found at the locations of some of the "woHII" GMCs ($R_{13/12} < 0.1$)
suggest that those "woHII" GMCs contain large fraction of diffuse cloud components and thus 
lower fraction of cloud cores compared to the "wHII" GMCs. 
\section {DISCUSSION}
%----------------------------------------
% Discussion. 1
%----------------------------------------
\subsection{Change of Cloud Property Across the Spiral Arm}
\label{sec: change_prop}
\begin {figure*} [htbp]
    %\plotone{img/cl_fig2_HII.ps}
	\plotone {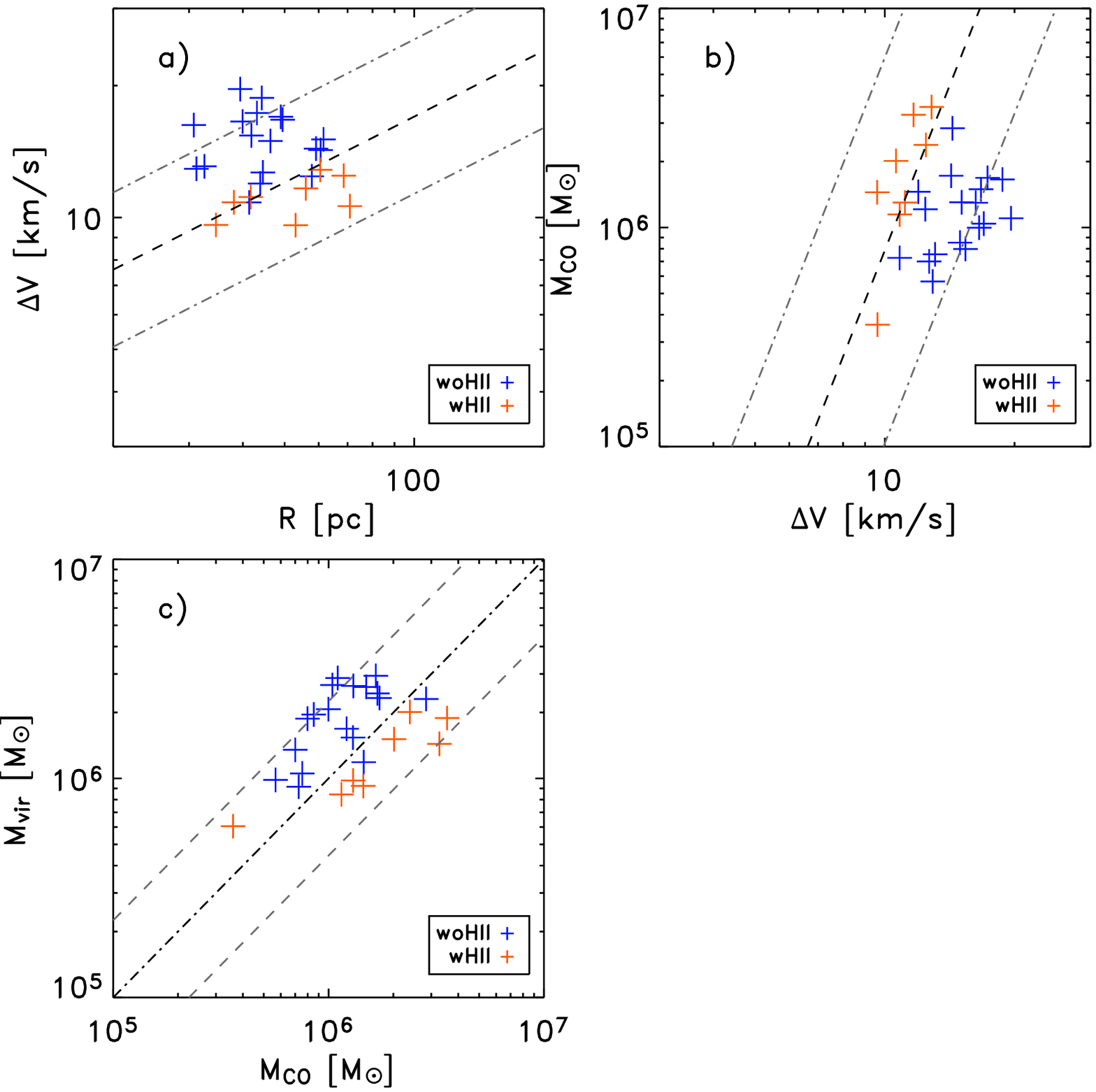}
	\caption {
    a) $\Delta{V} - R$ plot of the identified GMCs. 
    Red and blue crosses indicate the "wHII" GMCs and the "woHII" GMCs, respectively.
    Broken line indicates the scaling relation of \citet{Solomon1987Larson}.
    For comparison, the scaling relations with different coefficients (multiplied by 1.5 and 1/1.5, respectively) are 
    indicated with dash-dot lines.
    b) same as a), but for $M_{\rm{CO}} - {\Delta}V$.
    c) same as a), but for $M_{\rm{vir}} - M_{\rm{CO}}$.
    }
    % see 342arm07/memo_arm07.txt
	\label {fig: cl_fig2_HII}
\end {figure*}

\begin {figure} [htbp]
  \begin {center}
    %\plotone{img/hist_a12n_HII.ps}
	\plotone {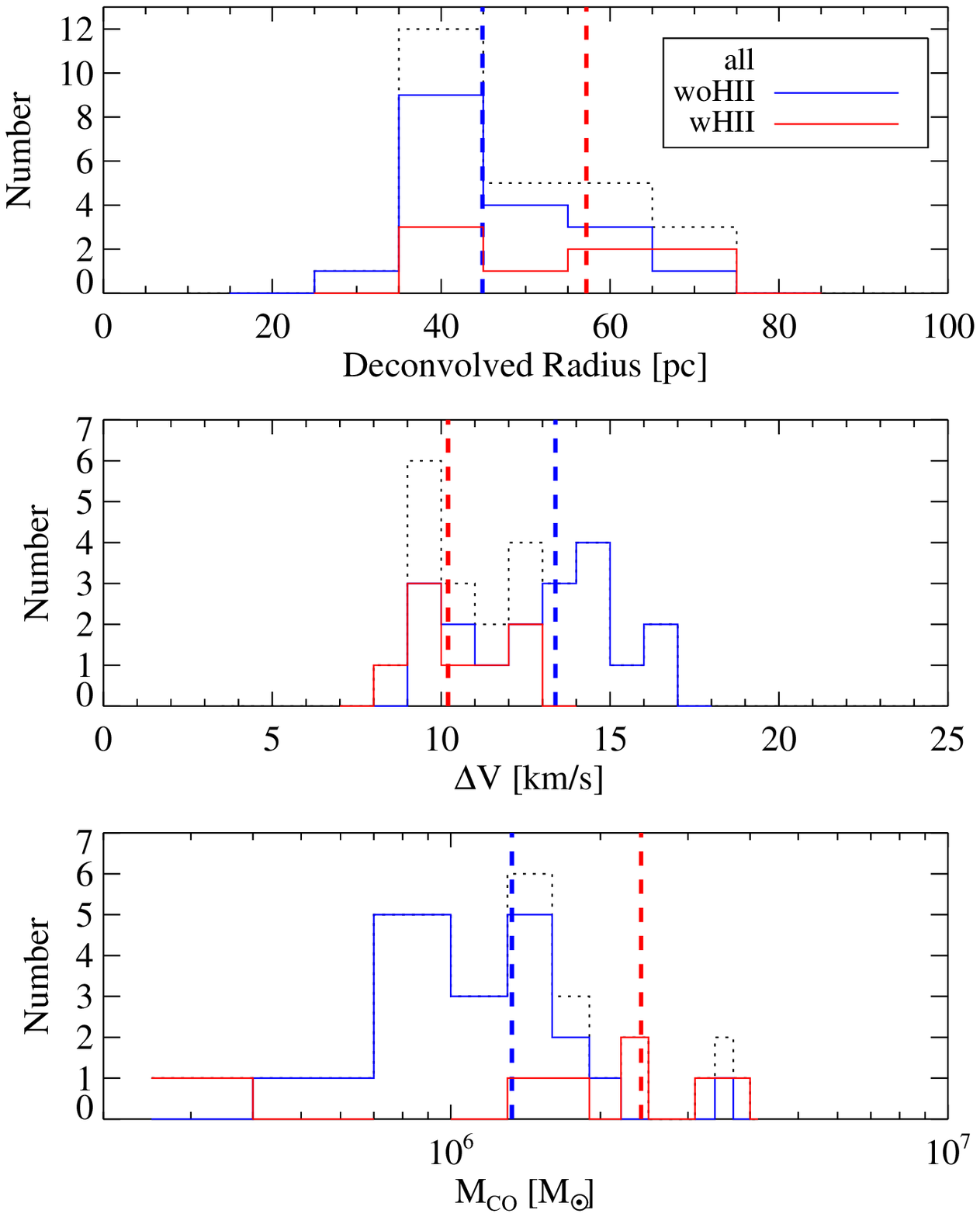}
    \caption {Histogram of the basic properties of the GMCs, namely, deconvolved effective radius, 
    deconvolved line width, and luminosity mass. 
    Histograms for the "wHII" GMCs, the "woHII" GMCs, and all GMCs 
    are indicated with red solid, blue solid, and dotted lines, respectively.
    Vertical dashed lines in red and blue indicate the median value of each category. }
    \label {fig: hist_a12_HII}
  \end {center}
% /home/hiko/reduction/342arm07/memo_arm07_robust0.txt
\end {figure}

\ \ The offset between the molecular gas traced by the $^{12}$CO(1--0) emission and massive star forming regions is often seen in grand-design spiral galaxies \citep[e.g.,][]{VogelKulkarniScoville1988, Rand1992m51COandHI, Rand1995AJNGC4321}.  Typical extent of the spatial offsets between the two components are typically $100$--$300$ pc. 
%However, resolutions of the previous observations toward spiral arms in the grand-design spiral galaxies were insufficient to investigate the variation of cloud properties within these offsets. 
However, most of the previous observations for such grand-design spiral galaxies, which are beyond the Local Group of galaxies, were at resolutions of $200$--$1000$ pc and were insufficient to investigate the variation of the molecular gas properties within these offsets.
The NMA+45m combined data provide a unique opportunity to look into the process involved there with a resolution of $\sim$ 50pc and an improved sensitivity to the extended diffuse emission.\\
\ \ Figure \ref{fig: cl_fig2_HII} shows scatter plots of the properties of clouds for each category.
%--------
% Panel1
%--------
The $R - \Delta{V}$ plot (Figure \ref{fig: cl_fig2_HII}(a)) shows that 
"woHII" GMCs have larger line width compared to the "wHII" GMCs with the same size. 
The Kolmogorov--Smirnov test was performed to the ratio between $R$ and $\Delta{V}$ with a null hypothesis of both categories being extracted from the same parent sample. The null hypothesis was rejected with a $p$-value of $\sim$ 0.004.
A similar trend is also seen in the $\Delta{V}$-$M_{\rm{CO}}$ plot (Figure \ref{fig: cl_fig2_HII}b).
Again, the Kolmogorov--Smirnov test was performed and indicated that there is a significant difference between the both categories with a $p$-value of $\sim$ 0.001.
Another difference of properties between the "wHII" and the "woHII" GMCs 
is seen in the $M_{\rm{CO}}$-$M_{\rm{vir}}$ plot (Figure \ref{fig: cl_fig2_HII}(c)). 
There is a tendency that the "woHII" GMCs have a larger virial mass to luminosity mass ratio than the "wHII" GMCs ($p$ $\sim$ 0.001), suggesting that "woHII" clouds are less gravitationally bound compared to the "wHII" clouds.\\
%-------------------------
% Prop/Histogram
%-------------------------
\ \ Figure \ref{fig: hist_a12_HII} shows histograms of the basic properties of the clouds (radius, line width, and mass).
While a little difference between the two categories is seen in the radius histogram,
apparent differences are seen in the mass and the line width histograms. 
As we have seen in the previous scatter plots, the line width distribution of the "woHII" GMCs seems to be different from the "wHII" GMCs ($p$ $\sim$ 0.017, for the Kolmogorov--Smirnov test).
The median line width of the "woHII" GMCs is $\sim$ 13.4 km s$^{-1}$ and 
is larger than that of the "wHII" GMCs ($\sim$ 10.2 km s$^{-1}$). 
Two cloud categories also show difference in the mass histogram ($p$ $\sim$ 0.011).
The mass of the "woHII" GMCs is concentrated around the median value of 
$ 1.3 \times 10^6 M_{\odot}$, 
except for the cloud-5 with the mass of $\sim$ 3.7 $\times 10^6$ $M_{\odot}$, 
which is located at the center of the GMA. 
On the contrary, the mass distribution of the "wHII" GMCs is more widely spread 
and stretched to a higher mass regime, up to 4.0 $ \times 10^6 M_{\odot}$.
Except for a cloud-26, all masses of "wHII" GMCs are larger than 
$10^6 M_{\odot}$. 
The cloud-26, which is located $\sim 300$ pc downstream from the molecular ridge,
has mass of 0.4 $ \times 10^6 M_{\odot}$ and is likely 
being disrupted by feedback from associating \ion{H}{2} regions.  \\
%-------------------------
% Prop change conclusion
%-------------------------
\ \ The comparison between the cloud properties between the two cloud groups
(the "wHII" and the "woHII" GMCs)
indicated significant variation of cloud properties 
according to their associated star forming activity. 
The "woHII" GMCs have larger line width and have smaller mass and moreover, 
are less gravitationally bound compared to the "wHII" GMCs.
Spatial distribution of the clouds indicate that 
the "wHII" GMCs are located downstream of the "woHII" GMCs, on average.
These facts suggest that 
cloud properties do change by crossing the spiral arm.
%Moreover, such variation of cloud properties may related to the onset of massive star formation.
%
%
%\ \ While mass of the "woHII" GMCs mainly ranges around $10^6 M_{\odot}$, 
%that of the "wHII" ranges from $10^6 M_{\odot}$ to $3.5 \times 10^6 M_{\odot}$. 
%One exception is cloud-26 with mass of $0.4 \times 10^6 M_{\odot}$, which is located 300 pc downstream from the molecular ridge. 
%The average mass of the "wHII" GMCs is $\sim 1.9 \times 10^6 M_{\odot}$ and clearly larger than that of the "woHII" ($1.2\times 10^6$ M$_{\odot}$). 
%The mean mass of the "woHII" GMCs is $\sim 1.2\times10^6$ M$_{\rm{\odot}}$ and is smaller than that of the "wHII" GMCs ($\sim 1.9\times10^6$ M$_{\rm{\odot}}$). \\
%This fact suggests that change of molecular gas properties across the spiral arm also involves the mass growth of the GMCs. 
%%
%
%
%\placefigure {fig: cl_fig2_HII}
%\placefigure {fig: hist_a12_HII}
\subsection {Effect of Gas Velocity Dispersion}
\label {sec: virial parameter}
\begin {figure} [htbp]
  \begin {center}
	%\plotone {img/cl_alpha.ps}
	\plotone {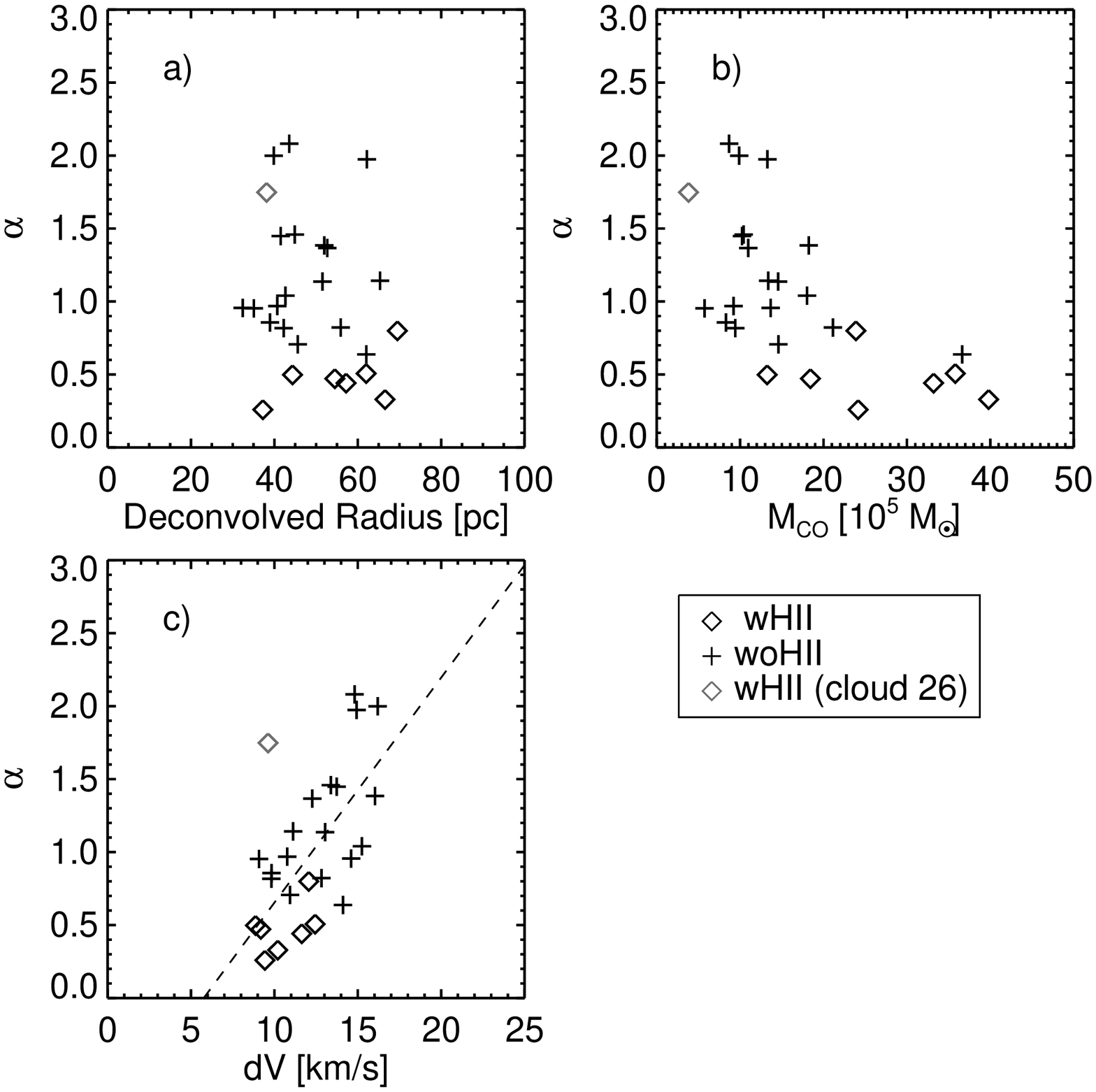}
    % scripts/batch_cl_alpha.pro
	\caption {
    (a) Relationship between the virial parameter
    ($\alpha_{\rm{vir}} (\equiv{M}_{\rm{vir}}/{M}_{\rm{CO}})$) 
    and radius ($R$).
    Crosses indicate the "wHII" GMCs and
    open triangles indicate the "woHII" GMCs. 
    A gray cross represents the cloud-26.
    (b) Same as (a), but for $\alpha_{\rm{vir}}$ and 
    $M_{\rm{CO}}$.
    (c) Same as (a), but for $\alpha_{\rm{vir}}$ and 
    $\Delta{V}$.
    Dashed line indicates a linear fit between the two quantities.
    }
	\label {fig: cl_alpha}
  \end {center}
\end {figure}

The comparison between the virial mass and the luminosity-based mass made 
in the previous subsection 
indicated that the "woHII" GMCs are 
loosely bound compared to the "wHII". 
The degree of gravitational binding can be expressed with a virial parameter,
\citep[$\alpha_{\rm{vir}}$:][]{BertoldiMcKee1992PressureMC}
, which is expressed by
%ratio of kinetic energy to the gravitational potential energy can be simply expressed as
\begin {equation}
    \alpha_{\rm{vir}} = M_{\rm{vir}} / M_{\rm{CO}}.
\end {equation}
As $\alpha_{\rm{vir}}$ decreases, cloud gets more unstable against gravitational collapse.
In this subsection, we will utilize $\alpha_{\rm{vir}}$ to investigate
which of the cloud properties is responsible for the observed variation of the 
gravitational instability of the clouds. \\
\ \ Figure \ref{fig: cl_alpha} shows the scatter plots between the 
basic cloud properties and $\alpha_{\rm{vir}}$. 
To see whether there exist a significant relation between the virial parameter $\alpha_{\rm{vir}}$ with other cloud properties, Kendall's $\tau$ and associated probability of chance correlation were calculated.
Significant correlation was only detected between the virial parameter $\alpha_{\rm{vir}}$ and the line width 
($p$ $\sim$ $6.0 \times 10^{-6}$).
GMCs in the Galactic center exhibit similar correlation between the 
line width and the degree of gravitational binding \citep{Oka2001GCMC}.
The correlation suggests that the degree of gravitational binding 
of the clouds is mainly determined by their internal turbulent motion. \\
\ \ The "wHII" GMCs have smaller velocity dispersion and $\alpha_{\rm{vir}}$, and
located downstream compared to the "woHII" GMCs.
Similar relation between the star formation activity and the line width was pointed out 
by \citet{Kohno1999NGC6951}.
They found an anticorrelation between the dense molecular gas fraction and 
the line width in the center of the barred galaxy NGC 6951, with a spatial resolution of $\sim 400$pc. 
The results presented here suggest that the anticorrelation between the star formation and the line width holds down to the size scale of GMC.
It is tempting to consider that dissipation of turbulent motion inside the GMCs 
occurs by crossing the spiral arm, and leads to the subsequent massive star formation.\\
%Ikuta \& Sofue (1997) found that the star formation efficiency (SFE) of the Milky Way GMC is inversely correlated with the velocity dispersion. \\
\ \ We must note that \citet{Scoville1986ClCollision} indicated the relation between the star formation activity and 
the line width, but in an opposite sense;
GMCs smaller than 40pc show that the GMCs associated with \ion{H}{2} regions have larger line width compared to the GMCs without \ion{H}{2} regions. 
The difference in the results is likely due to the difference in the observed size scales. 
Since the spatial resolution of the data used by \citet{Scoville1986ClCollision} is finer than this work (down to pc scale), the effects of local phenomena, 
such as shock induced by cloud--cloud collision and radiation feed backs from \ion{H}{2} regions, might be dominant in the previous result.
While, on the other hand, the resolution of this study is 
comparable to the size of the GMCs and 
the derived properties should be reflecting the 'global' properties of 
clouds rather than the internal phenomena within each cloud.
\subsection {Evolution of GMC in the Spiral Arm}
\label {sec: EvolutionOfGmcInTheSpiralArm}
\ \ The results presented in the previous subsections illustrated 
the variations of the cloud properties across the spiral arm in \objectname{IC 342}. 
The "wHII" (downstream) GMCs are more virialized and massive compared to the "woHII" (upstream) GMCs. 
The cloud groups also differ from each other in that the velocity dispersion of the "wHII" GMCs is larger than that of the "woHII" GMCs.
In this subsection, we will discuss the implications of such variations of the cloud properties. \\
\ \ First question is from what kind of material the clouds identified in this study were made?
There are two classes of concepts about the formation of GMCs in the spiral arm. 
One is that GMCs are formed from \ion{H}{1} clouds in the inter-arm 
and another is that GMCs are built-up from smaller clouds in the inter-arm through agglomeration.
As we have seen before in Suction \ref{sec: 342HI},
the surface mass density of \ion{H}{1} is only $\sim 2 M_{\odot}$ pc$^{-2}$ within the observed region. 
On the other hand, the surface mass density of the molecular gas is estimated to be 
$\sim 30 M_{\odot}$ pc$^{-2}$ from the single-dish $^{12}$CO data.
So, it is natural to consider that the GMCs in the spiral arm are formed from the pre-existing molecular clouds. \\
\ \ As the extent of the observed region is limited, 
we are unable to tell the nature of the inter-arm molecular clouds from the observed data alone. 
Instead, we refer to the observations of molecular clouds in the Milky Way. 
It is known that GMC with mass above $10^5$ $M_{\odot}$ is rarely seen in the inter-arm region 
\citep[e.g.,][]{SandersScovilleSolomon1985}. 
\citet{Heyer1998OuterGalaxy} indicated that  
inter-arm molecular clouds are smaller than $3\times10^4 M_{\odot}$ and not virialized 
in the outer Galaxy. 
Even in the molecular-rich inner Galaxy, there are substantial contributions of diffuse molecular emission, suggesting the existence of small clouds. \citep{Polk1988,Chiar1994CO21Survey}. 
It seems natural to assume that inter-arm molecular clouds are dominated by 
the small molecular clouds. \\
%
%\ \ The theories of GMC formation claim that show that it is possible to build up GMC as massive as $10^6 M_{\rm{\odot}}$ from the small clouds in the inter-arm. 
%
%
%
%
\ \ Second question is what kind of mechanism could explain the observed variations the of cloud properties?
If the inter-arm clouds in \objectname{IC 342} are small, diffuse clouds, the mechanism for building up the GMCs as massive as $10^6 M_{\rm{\odot}}$ must exist. 
The timescale for building up should be comparable to the arm crossing time
in this region 
\citep[$t_{\rm{arm}} \sim 3\times10^7$yr,][]{Hirota2010IC34213CO}.
The random coalescence of molecular clouds is proposed as a formation mechanism of GMC 
\citep[][]{KwanValdes1983MassGrowth, CasoliCombes1982CloudFormationInArm, Tomisaka1984ClCoagulation, RobertsStewart1987CloudCloudCollision, Elmegreen1990MCFormationTheories, Dobbs2008AgglomerationSelfGravity}. 
Early calculations indicated that it may take about $10^8$ yr, 
which is far longer than $t_{\rm{arm}}$.
%In most case of the calculations, clouds are treated as sticky particle and 
%coalesce or disrupt when they interact each other, depending on the impact parameter. 
However, if the spiral density waves are included, 
it turns out that the formation of the GMC as massive as several times $10^6 M_{\odot}$ within a time comparable with $t_{\rm{arm}}$ is possible.
Large-scale instability induced by the self-gravity of gas shall further make the agglomeration process efficient
\citep[e.g.,][]{BalbusCowie1985GravitationalInstability, Dobbs2008AgglomerationSelfGravity}.\\
\ \ If the random coalescence model is adopted, 
the observed nature of the "wHII" and the "woHII" GMCs could be understood naturally.
GMCs were made from the small molecular cloud in the inter-arm through agglomeration.
The end product of the cloud growth will be the "wHII"("post arm") GMCs, 
which have masses of (1--3) $\times 10^6 M_{\odot}$ (an exception is the cloud-26). 
As the masses of the "woHII" GMCs are around $10^6 M_{\odot}$ and smaller compared to the "wHII" GMCs, 
it may be natural to consider that 
the "woHII" GMCs are at the intermediate stage of the cloud growth by coalescence.
The relatively low peak temperatures of the "woHII" GMCs ($T_{\rm{MB}}$ = 2--3K) and the low degree of gravitational binding ($\alpha_{\rm{vir}}$ = 1--3) are in agreement with the notion that 
the "woHII" GMCs are collective of smaller clouds, 
which are yet not virialized.\\
\ \ Above interpretation may be in agreement with the 
recent results of observations of the grand-design spiral galaxy M51.
\citet{Koda2009M51} performed full aperture observation of M51 in $^{12}$CO (1--0)
with a spatial resolution of $\sim$ 200 pc and indicated 
that while massive GMAs are preferentially found in the spiral arm, 
GMCs with mass below 10$^6$ $M_{\odot}$ are uniformly distributed over the molecular gas disk.
\citet{Egusa2011M51} carried out CO observations toward the selected region of M51 with a spatial resolution of 30pc, comparable with ours, but not corrected for missing-flux, 
and found that clumps with mass of 10$^{5}$--10$^6$ $M_{\odot}$ are likely preferentially located on the downstream side of the spiral arm. 
The combined data presented here revealed while rather diffuse "woHII" clouds are the present at the upstream side, discrete clouds are seen on the downstream side. 
While the buildup of GMAs by coagulation of GMCs in the inter-arm of M51 was suggested by \citet{Koda2009M51}, 
the low $R_{13/12}$ in the center of the GMA and presence of diffuse clouds in the upstream of the spiral arm implies coalescence of diffuse clouds are much important in the arm of \objectname{IC 342}.\\
%--------------------------------------
%In this case, the relatively larger velocity dispersions of the "woHII" GMCs are also naturally explained.
%
%
%
%
\ \ Another important aspect of the variation of cloud properties is the decrease of line width.
%Relatively large velocity width can be understood since the observed line width of the "pre arm" GMCs are combination of the line width of small clouds and cloud-cloud velocity dispersion between small clouds. \\
%
% Decrease of line width via inelastic collision
%
%\ \  Are there any other mechanism for dissipating the line width of the GMCs?.
%In fact, the variation of velocity dispersion also may be explained with the collisional coalescence model. 
Since molecular clouds collide inelastically, part of the kinetic energy should be transformed into thermal energy and radiated. 
\citet{Tomisaka1987CloudFluidGalacticShock} carried out the numerical simulation of collisional clouds 
incorporating formation and destruction of GMCs, the loss of random velocity due to inelastic cloud collision, and the energy input by the star formation within GMCs. 
The results were that after the compression of cloud by the shock, 
random velocity of clouds decreases and density increases. 
These are in agreement with the results presented in previous subsections. \\
\ \ The scenario of cloud growth would be summarized as follows. 
First, the inter-arm molecular clouds are collected because of the convergence of stream line before the spiral arm. 
Due to this convergence of stream line and self-gravity, clouds will stick and form the cloud collective with masses of $10^6 M_{\odot}$, 
although they are not fully virialized yet and loosely bound. 
These cloud collectives are observed as the "woHII" GMCs. 
Once massive GMC is formed, its large gravitational cross section accelerates the coalescence and swallows diffuse clouds at the envelope. 
The excess kinetic energy will be dissipated due to inelastic collision and virialized GMCs with mass larger than $10^6 M_{\odot}$ are formed (the "wHII" GMCs").
\subsection {Condition of Massive Star Formation}
\begin {figure} [htbp]
  \begin {center}
	%\plotone {img/a12_alpha_flux.eps}
	\plotone {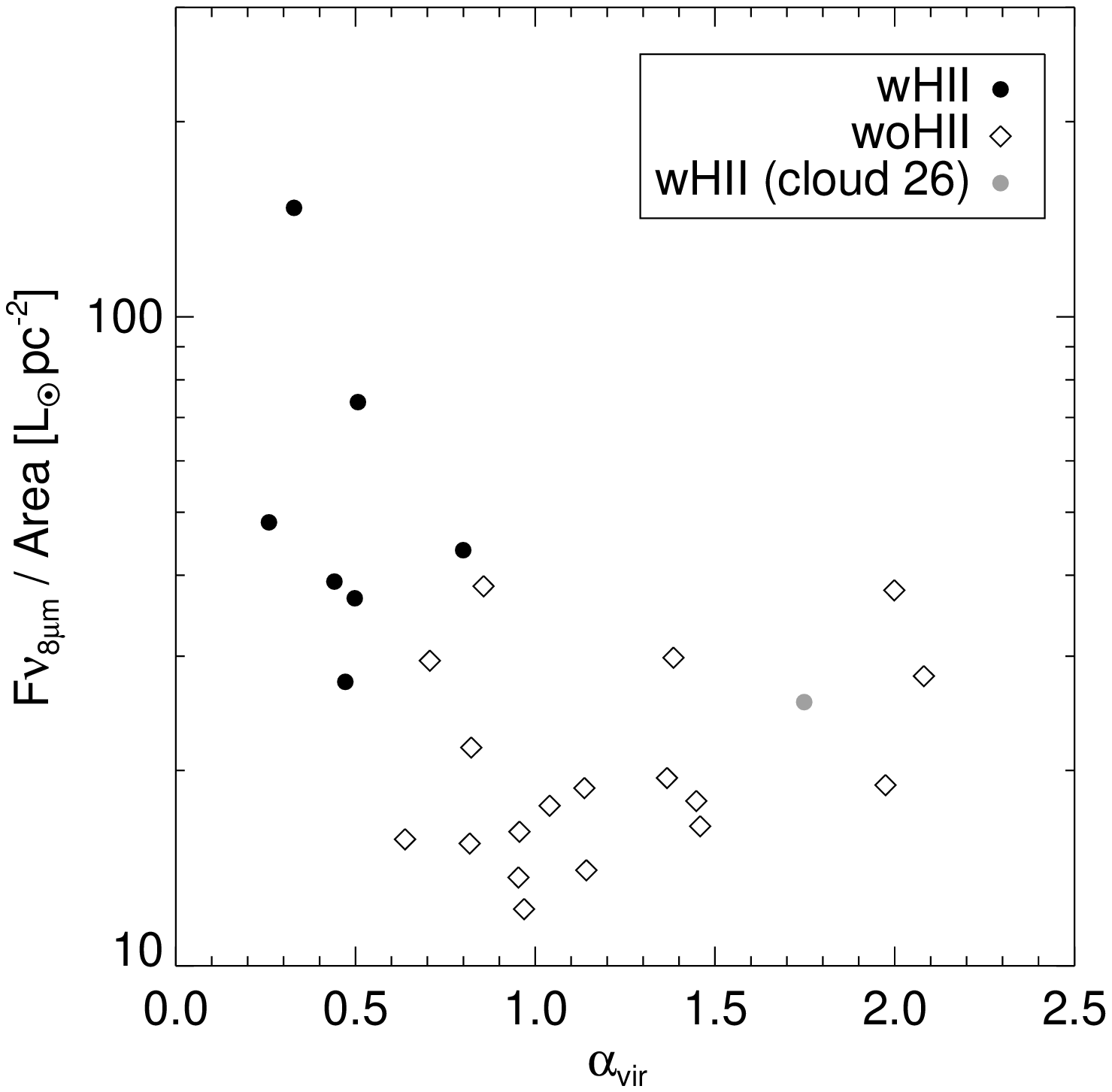}
	\caption {Scatter plot of area-averaged 8$\mu$m luminosity measured within each cloud boundary and 
    the viral parameter ($\alpha_{\rm{vir}}$) for each cloud.
    Filled circles indicate the "wHII" GMCs, open diamonds the "woHII" GMCs, and filled gray circle indicates the cloud-26. }
	\label {fig: a12_alpha_flux}
  \end {center}
% /home/hiko/temp/temp.pro
\end {figure}

\if0
    \ \ Without any support, GMC should collapse on its free-fall time, $t_{\rm{ff}}\sim4$ Myr. 
    If this is the case, the SFR in the Milky Way galaxy which contains $10^9 M_{\odot}$ of 
    molecular gas mass is $\sim 250 M_{\odot}$yr$^{-1}$. 
    However, the exact SFR observed in the Milky Way is only $\sim 3 M_{\odot}$yr$^{-1}$ (McKee \& Williams 1997). 
    This low SFR is one of the major unresolved problems in the theories of the ISM. \\
    \ \ There are two ways for the solution of this problem. 
    One is that setting the time scale for star formation ($t_{\rm{SF}}$) longer, 
    and another is lowering the star formation efficiency per GMC per unit time ($\epsilon_{\rm{SF}}$). 
    The star formation efficiency of GMC (SFE$_{\rm{GMC}}$) is expressed by, 
    \begin {equation}
    SFE_{\rm{GMC}} = t_{\rm{sf}} * \epsilon_{\rm{SF}} \sim 0.01
    \end {equation}
    where $t_{\rm{sf}}$ is the time scale of star formation, and $\epsilon_{\rm{SF}}$ is the star formation rate per unit time. \\
\fi
%The CO-H$\alpha$ offsets seen in M51 is considered as the evidence of massive star formation. The time offsets estimated from the spatial offsets was $1-3 \times 10^7$ yr, which is close to the Jeans time scale. It is considered that gravitational collapse of GMA 
\ \ The closeness between the star forming regions and the "wHII" GMCs suggests that
these clouds are real progenitors of the associated \ion{H}{2} regions.
This can also be justified by the consideration of the age of \ion{H}{2} regions. 
Sizes of the \ion{H}{2} regions seen in the H$\alpha$ image have at most a diameter of $\sim 100$ pc. 
Assuming sound speed in the ionized gas as $c \sim 10$ km s$^{-1}$, 
an upper limit on the age of the \ion{H}{2} regions could be estimated as $D/2c = 5$ Myr. 
As the age of \ion{H}{2} regions estimated is much shorter than the arm crossing time
($t_{\rm{arm}} \sim 3\times10^7 M_{\odot}$, ), 
the \ion{H}{2} regions must have been born within the associated "wHII" GMCs on site.
The relatively young age of the \ion{H}{2} regions compared to $t_{\rm{arm}}$ 
gives us another important implication; 
if the "woHII" GMCs are able to initiate massive star formation, 
we should see the symptoms of massive star formation inside the "woHII" GMCs.
Nevertheless, little 8$\mu$m emission could be seen in the "woHII" GMCs. 
This implies that there must be a condition for massive star formation, 
which is fulfilled by the "wHII" GMCs but not by the "woHII" GMCs. \\
\ \ As we have seen in Section \ref{sec: EvolutionOfGmcInTheSpiralArm}, 
collisional coagulation of clouds seems to be responsible for the evolution of the GMCs. 
One of the possible mechanism for initiating massive star formation is 
cloud--cloud collision \citep{Scoville1986ClCollision, Tan2000ClCollisionStarburst}.
In this picture,  molecular clouds are compressed by the shock at the colliding surface 
and subsequently dense cores are formed \citep{Kimura1996ClCollision}. 
However, at the center of the GMA where large velocity gradient due to streaming motions is observed 
and enhanced rate of collision is expected, little or no sign of star formation is seen. 
This implies that for the onset of massive star formation, 
not only cloud collision but also some other conditions may be required. 
Another interpretation as following might arise.
Namely, after star formation is initiated inside the "woHII" clouds, 
star formation proceeds rather slowly and become only visible at the "wHII" clouds. 
However, this requires slow progress of star formation, with a timescale comparable 
to the arm-crossing time mentioned above.
Even if this is the case, the progress of the star formation shall be regulated by the evolution of GMCs 
because of the far longer timescale of arm-crossing time compared to dynamical timescale in each GMC
\citep[$\sim4\times10^6$yr, e.g., ][]{Pringle2001}. 
Both interpretations requires the evolution of the cloud properties which enable triggering of star formation
to work effectively.  \\
\ \ Figure \ref{fig: a12_alpha_flux} shows a scatter plot of the area-averaged 8$\mu$m luminosity 
measured within each cloud boundary and the virial parameter $\alpha_{\rm{vir}}$ for each cloud.
There is a tendency that 8$\mu$m luminosity, which is proportional to the star formation rate, 
is higher in the range of $\alpha_{\rm{vir}} < 1$ compared to in the range of $\alpha_{\rm{vir}} > 1$. 
It was shown in Section \ref{sec: virial parameter} that the degree of binding of the GMCs is mainly determined by its line width. 
These facts suggest that virialization of GMC achieved by the dissipation of 
turbulent motion is one of the necessary conditions for the onset of massive star formation. \\
\ \ Recent studies emphasize the importance of turbulence in both regulating and the triggering star formation. 
In the theory of turbulence regulated star formation, although dense cores are formed by the shock driven by supersonic turbulence inside a GMC
\citep[e.g.,][]{Padoan2002TurbulentFragmentation2CloudCore, MacLow2004TurbulenceRegulatedSF}, limited part of the GMC can be dense enough to form cloud cores since the greater part of the GMC are supported by the excess kinetic energy \citep{Elmegreen2002SFOverCriticalDensity}.
%only a limited fraction of the GMC can be dense enough to form cloud cores since 
%the greater part of the GMCs are supported by the excess kinetic energy \citep{Elmegreen2002SFOverCriticalDensity}.
\citet{Krumholz2005TurbulenceSF} presented semianalytical calculation using the theory and showed that amount of the excess kinetic energy inside the GMC, which is proportional to $\alpha_{\rm{vir}}$, controls the production rate of cloud cores, and thus star formation rate. 
This point is in agreement with our results that dissipation of turbulence is required for the onset of massive star formation. 
After GMCs have dissipated excess kinetic energy, it can initiate star formation, whatever the trigger may be. 

\section {SUMMARY}
Results of the full aperture observations of the northeastern spiral arm segments of \objectname{IC 342} in $^{12}$CO (1--0) and $^{13}$CO (1--0) made with a spatial resolution of $\sim$ 50 pc were presented. 
Though the spatial resolution of the observations is not high enough to resolve internal structure of GMCs, 
it is capable of investigating the global property of massive GMCs. 
The observations cover the 1kpc $\times$ 1.5kpc region which contain a GMA with mass of $\sim 10^7$ $M_{\odot}$, where spatial offsets between $^{12}$CO and star-forming regions exist. 
The NMA data were combined with the 45m telescope data to 
image the extended CO distribution and 
investigate whether and how the properties of molecular clouds change by crossing the spiral arm.
The results are summarized as follows.
\begin {enumerate}
%\item The NRO 45m observation of the IC 342 in $^{13}$CO (1--0) line was executed in OTF observation mode. The OTF observation successfully mapped the $^{13}$CO (1--0) line with sufficient sensitivity and resolution to resolve galactic structures of IC 342.
\item 
The spiral arms were resolved into a number of clouds which have size, temperature, and surface mass density comparable to the massive GMCs in the Galaxy. 
While the $^{12}$CO (1--0) distribution is concentrated in a narrow ridge in the southern half of the observation field, it is more smoothly distributed in the northern half, where the GMA exists. 
Comparison with the $^{13}$CO (1--0) image showed that at the center of the GMA only little $^{13}$CO (1--0) emission is detected despite of its strong $^{12}$CO emission. 
\item Comparison with star formation tracers indicated that 
while some of the clouds are closely associated with star forming regions, 
the rest of the clouds show little or no sign of massive star formation. 
The identified clouds were divided into two categories according to whether 
they are associated with star formation activity or not.
%On the other hand, discrete sources located at the southern half of the observed region are closely associated with HII regions.
%   \item Kinematics of the molecular gas shows velocity shift on the trailing side of the spiral arm. 
%   The degree of velocity shift is $\sim$ $300 $km s$^{-1}$ kpc$^{-1}$, much smaller than found in the grand design spiral galaxy \objectname{M51}. 
%   The gas motion is interpreted as moving radially outward on the trailing side of the spiral arm and changes its direction on the leading side, streaming along the spiral arm. 
\item Twenty-sixclouds with masses of ($0.4$--$3.7$) $\times$ $10^6$ $M_{\odot}$ were identified from the $^{12}$CO (1--0) data cube.
The size and the line width of the clouds were comparable to GMCs in the Milky Way. 
The identified clouds also followed the linewidth--size relation and 
the mass--linewidth relation of the Galactic GMCs, suggesting that the identified clouds are resemblance of Galactic GMCs. 
\item The identified clouds (GMCs) were divided into two categories according to whether they are associated with star-forming regions or not. 
Comparison between both categories showed that clouds which are associated 
with star-forming regions ("wHII" GMCs) are more 
virialized and massive compared to the clouds which show little or no sign of star formation ("woHII" GMCs).
Moreover, the "woHII" GMCs have larger line width compared to the "wHII GMCs". 
As the "wHII" GMCs are located downstream of the "woHII" GMCs, 
it is concluded that properties of the GMCs do change by crossing the spiral arm. 
\item The line ratio, $R_{13/12}$, varies from $0.06 \pm 0.01$ in the center of the GMA where little or 
no sign of star formation is seen to $\sim 0.16$ in the discrete GMCs which are associated with \ion{H}{2} regions. 
On the whole, the "woHII" GMCs likely show a low $R_{13/12}$ value which suggests a larger fraction of diffuse molecular cloud components than the "wHII" GMCs.
\item To see what parameter of the GMC controls degree of gravitational boundness of the GMC, 
correlation plots between the virial parameter, $\alpha_{\rm{vir}} = M_{\rm{vir}} / M_{\rm{CO}}$, and other basic GMC properties, namely, radius, line width, and mass were made. The linewidth--$\alpha_{\rm{vir}}$ plot indicated most strong correlation implying that the dissipation of turbulent motion controls the boundness of the GMCs. 
\item Random coagulation of the pre-existing small clouds can explain the growth of the mass and the timescale involved in the change. Collisional coalescence involved in random coagulation may also explain the observed decrease of line width, as the kinetic energy of the clouds will be dissipated by inelastic cloud collision. 
\item 
The closeness of \ion{H}{2} regions and the "wHII" GMCs imply that those GMCs are real progenitors of the associated \ion{H}{2} regions. 
The rough estimate of the age of \ion{H}{2} regions gives $\sim$ 5 Myr, which is well below the arm-crossing time. 
This implies if the "woHII" GMCs are able to initiate star formation, the symptoms should have seen. So, the sparseness of star formation in the "woHII" GMCs suggests they do not meet the condition for massive star formation. 

Plot of area-averaged 8$\mu$m luminosity and virial parameter ($\alpha_{\rm{vir}}$) shows a tendency that the 8$\mu$m luminosity, which is proportional to the star formation rate, rises in the range of $\alpha_{\rm{vir}} < 1$.
This might imply that dissipation of turbulence controls not only the evolution of GMCs, but also the probability of massive star formation. 
\end {enumerate} 
\acknowledgments
The authors would like to thank the Nobeyama Radio Observatory (NRO) staff for operating the NMA and the 45m telescope. NRO is a division of the National Astronomical Observatory of Japan under the National Institutes of Natural Science. 
%This is an acknowledgment.

% We are grateful to V. Barger, T. Han, and R. J. N. Phillips for
% doing the math in section~\ref{bozomath}.
% More information on the AASTeX macros package is available \\ at
% \url{http://www.aas.org/publications/aastex}.
% For technical support, please write to
% \email{aastex-help@aas.org}.


%% in parentheses, after the keyword, but they will not be verified.
%{\it Facilities:} \facility{NRO}.
%{\it Facilities:} \facility{Nickel}, \facility{HST (STIS)}, \facility{CXO (ASIS)}.
{\it Facilities:} \facility{No:45m}, \facility{NoMA ()}
\end {document}